\documentclass{emulateapj}
\usepackage{graphics}

\newlength{\columnfigurewidth}
\newlength{\twocolumnfigurewidth}
\begin{document}

\title{The isotropic diffusion source approximation for supernova neutrino
transport}

\author{M. Liebend\"orfer, S. C. Whitehouse, T. Fischer}
\affil{Department of Physics, University of Basel, Klingelbergstr. 82,
CH-4056 Basel, Switzerland}

\begin{abstract}
Astrophysical observations originate from matter that interacts
with radiation or transported particles. We develop a pragmatic approximation
in order to enable multi-dimensional simulations with basic spectral radiative
transfer when the available computational resources are not sufficient to solve
the complete Boltzmann transport equation. The distribution function
of the transported particles is decomposed into a trapped particle
component and a streaming particle component. Their separate evolution
equations are coupled by a source term that converts trapped particles
into streaming particles. We determine this source term by requiring
the correct diffusion limit for the evolution of the trapped particle
component. For a smooth transition to the free streaming regime, this
'diffusion source' is limited by the matter emissivity. The resulting
streaming particle emission rates are integrated over space to obtain
the streaming particle flux. Finally, a geometric estimate of the
flux factor is used to convert the particle flux to the streaming
particle density, which enters the evaluation of streaming particle-matter
interactions. The efficiency of the scheme results from the freedom
to use different approximations for each particle component. In supernovae,
for example, reactions with trapped particles on fast time scales
establish equilibria that reduce the number of primitive variables
required to evolve the trapped particle component. On the other hand,
a stationary-state approximation considerably facilitates the treatment
of the streaming particle component. Different approximations may
apply in applications to stellar atmospheres, star formation, or cosmological
radiative transfer. We compare the isotropic diffusion source approximation
with Boltzmann neutrino transport of electron flavour neutrinos in
spherically symmetric supernova models and find good agreement. An
extension of the scheme to the multi-dimensional case is also discussed.
\end{abstract}
\keywords{supernovae: general---neutrinos---radiative transfer---hydrodynamics--
-methods: numerical}

\section{Introduction}

Most applications in computational astrophysics involve matter that
is in thermal or reactive equilibrium. Astrophysical observations,
however, require information to propagate from the astrophysical site
to a terrestrial observer. Hence, the observationally interesting
events involve additional particle species that are not trapped in the fluid and
must therefore be treated by radiative transfer. In many traditional
numerical models, the equilibrated matter is treated in the hydrodynamic
limit, while transported particle species are treated by a suitable
algorithm of radiative transfer. If this splitting in the numerical
method is based on the particle species, it can become extremely challenging
and inefficient if the range of conditions in the astrophysical scenario
is large. In situations where radiative particles are in thermal
or reactive equilibrium, the non-local algorithm of radiative transfer
has to be perfectly consistent with the hydrodynamics scheme and capable
of handling very stiff source terms in order to evolve the applicable
equilibria.

For example core collapse supernova events have challenged computer
models for several decades \citep{Colgate.White:1966,Arnett:1966,Bowers.Wilson:1982,Bethe.Wilson:1985,Bruenn:1985,Myra.Bludman:1989}.
These stellar explosions occur at the end of the life of a massive
star when the growing iron core, as the end product of nuclear fusion,
becomes unstable against gravitational collapse. The collapse is halted
only after nuclear density is reached at the centre. The repulsive
strong interaction and the neutron degeneracy pressure reduce the
compressibility of matter so that a sound wave travels outward
through the causally connected inner core. It turns into a shock wave
when it reaches supersonically accreting matter from the outer layers
and, due to dissociation and neutrino losses, stalls within few milliseconds
to an initially hydrostatic accretion front, which is thought to slowly
expand over several hundreds of milliseconds before the explosion
sets in. The new-born protoneutron star
(PNS) at the centre of the event has a larger density by many orders
of magnitude than the shock-heated and dissociated matter that continues
to accumulate behind the accretion front. This accreted matter is
subject to a variety of fluid instabilities and asymmetric flow patterns
that couple to the energetic neutrino radiation field produced within
or close to the surface of the PNS. The emergence of an explosion
and the detailed interaction between the multi-dimensional flow and
the neutrino field is subject to a long-standing and ongoing debate.
See for example \citep{Bethe:1990} for a theoretical description of
the scenario or \citep{Bruenn.Hix.ea:2006,Dessart.Burrows.ea:2007,Marek.Janka:2007}
for recent axisymmetric computer simulations with spectral neutrino
transport.

Several aspects of the core collapse supernova scenario make computer models
difficult to perform: The relevant densities in the computational domain range
over ten decades from about \( 10^{15} \) g/cm\( ^{3} \) at the
centre of the PNS down to about \( 10^{5} \) g/cm\( ^{3} \) in the
outermost layers (the star is much larger but does not dynamically
react to the changes at the centre during the first few hundred milliseconds
after bounce). This density range implies very disparate dynamical
time scales. At high densities, the weak interaction rates are much
faster than the dynamical time scale, while at low density, they are
even slower than the evolution time scale. This requires implicit
finite differencing of many processes to perform simulations in a
reasonable amount of time. Moreover, the gravitational
and stored internal energy scales are of order \( 10^{53} \) erg,
while typical kinetic explosion energies are two orders of magnitude
smaller. The presence of a strong shock
with near-relativistic infall velocities further challenges the resolution
and stability of numerical models. In this very dynamical environment
with multi-dimensional fluid instabilities and differential rotation,
a computer model has to quantify the exchange of energy between matter
and the emitted neutrino field in order to allow statements about
the explosion dynamics to be made. As the neutrino cross sections at given matter
conditions scale with the square of the neutrino energy, the neutrino
transport problem must be considered as a superposition of several
single-energy (monochromatic) transport problems, where the geometry of
the scattering sphere depends on the neutrino energy and type.
This leads to multi-group transport schemes where
all propagation quantities are solved in separate energy groups \citep{Mazurek:1975,Arnett:1977}.
Moreover, stationary-state studies with spherically symmetric Boltzmann
neutrino transport showed that the angular distribution of the neutrino
propagation directions have also to be taken into acount to obtain
accurate neutrino heating and cooling rates \citep{Wilson:1971,Mezzacappa.Bruenn:1993a,Messer.Mezzacappa.ea:1998,Yamada.Janka.ea:1999,Burrows.Young.ea:2000}.

In the mean time, the spherically symmetric Boltzmann transport equation
is routinely solved for all neutrino flavours in dynamical models
of stellar core collapse and postbounce evolution \citep{Liebendoerfer.Mezzacappa.ea:2001,Rampp.Janka:2002,Thompson.Burrows.Pinto:2003,Sumiyoshi.Yamada.ea:2005}.
These comprehensive and sophisticated codes treat disparate time scales by implicit
finite differencing and require large computational resources. However,
due to the enormous density range in the scenario, a considerable
fraction of the computational domain is either neutrino-opaque or
neutrino-transparent. In a fully neutrino-opaque regime, a numerical
solution to the Boltzmann transport equation is rarely more accurate
than the solution of the simpler diffusion equation that describes
the correct physical limit. Similarly, the angular distribution of
the neutrino propagation directions in the far-field free streaming
regime is precisely determined by the geometry and emissivity of the
distant sources, while an angular discretisation of the Boltzmann
equation may lead to discretisation errors. Hence, we suggest that
a combination of different approaches to radiative transfer is applied
in the same astrophysical simulation. This leads to the concept of an
'adaptive algorithm', which should not be regarded as an inconsistency
or drawback. On the contrary, it saves computation
power that can be invested for better resolution, more input physics,
or the coverage of a larger parameter space. Current state-of-the-art
supernova models with spectral neutrino transport are performed under
the assumption of axisymmetry. As a 2D solution of the Boltzmann equation
is computationally inefficient for the reasons discussed above, different
groups either apply 1D Boltzmann solutions in separate angular segments
in combination with 2D neutrino advection in opaque regimes (ray-by-ray)
\citep{Marek.Janka:2007}, resort to the multi-group flux-limited diffusion
(MGFLD) approximation \citep{Walder.Burrows.ea:2004,Ott.Burrows.ea:2008}, or
use a combination of both \citep{Bruenn.Hix.ea:2006}. 

Three-dimensional supernova models were hitherto only affordable with
'grey' neutrino transport \citep{Fryer.Warren:2004} or with other
simple ad hoc approximations to the neutrino physics \citep{Scheck.Plewa.ea:2004,Ott.Dimmelmeier.ea:2007,Scheidegger.Fischer.ea:2008}.
But many features of the supernova should be studied in three spatial
dimensions, for example the pattern of the accretion flow \citep{Herant.Benz.ea:1994},
the standing accretion shock instability \citep{Blondin.Mezzacappa.DeMarino:2003},
the excitation of PNS g-modes \citep{Burrows.Livne.ea:2006}, and the
evolution of magnetic fields (see e.g. \citet{Kotake.Sato.Takahashi:2006}
and references therein). In order to enable three-dimensional supernova
models with spectral neutrino transport and in order to accelerate
parameter studies with simulations of lower dimensionality, we construct
the isotropic diffusion source approximation based on several
ideas scattered across the literature. Our algorithm is not meant to compete
in accuracy with more detailed solutions of the transport equation in the semi-transparent
regime. As the transition from the opaque to the transparent regime will be handled
by interpolation and basic geometric considerations, it is important that the approach
is verified by an accurate transport solution for every new field of application. Our
approach shares this limitation with the well-known flux-limited diffusion approximation.
The main advantage of the isotropic diffusion source approximation over the latter is that the
fluxes and flux factors in the transparent regime are determined by the non-local distribution
of sources rather than the local intensity gradient, which may give an incorrect flux direction.
Moreover, the new approach tries to limit the computationally expensive solution of the multi-dimensional diffusion equation to the opaque regime where the diffusion approximation
is adequate. The goal is to create a flexible algorithm that efficiently implements only the
dominant features of radiative transfer in one consistent framework.

The most obvious method of avoiding the inefficient global application
of algorithms is a decomposition of the problem in space, so that
one algorithm is used in one subdomain while another algorithm is
used in another subdomain \citep{Chick.Cassen:1997}. Another approach
is the decomposition in the momentum phase space into particles with
thermal velocities and particles with supra-thermal velocities. If
one treats the thermal particles by an efficient fluid model, a hybrid
kinetic/fluid model is obtained \citep{Crouseilles.Degond.Lemou:2004}.
The isotropic diffusion source approximation described in this article
is more similar to the so-called \( \delta f \)-method \citep{Parker.Lee:1993,Brunner.Valeo.Krommes:1999}.
We also decompose the distribution function of the transported particles
into a thermal component and a perturbation that is allowed to overlap
the thermal part in the entire computational domain and particle phase
space. However, in our case, the perturbation is not considered to
be small. Inspired by flux-limited diffusion \citep{Levermore.Pomraning:1981},
the evolution of the radiation component is guided by the diffusion
limit at large opacities and by the free streaming limit at low opacities.
In contrast to flux-limited diffusion, we build our scheme conceptually
on approximations of the collision integral rather than particle fluxes.
We assume that the free streaming particle flux is dominated by the
flux emerging
from the diffusive domain so that the sources for the far field can
easily be integrated. In the stationary-state limit, the determination
of the particle fluxes reduces to the solution of a Poisson equation
\citep{Gnedin.Abel:2001}. Since the matter interacts according to
the local particle abundances and not fluxes, we have to convert the
particle fluxes to local particle densities. This is achieved by a
geometric estimate of the flux factor as suggested and evaluated by
Bruenn in \citep{Liebendoerfer.Messer.ea:2004}.

In Sect. \ref{sec:idsa} we describe in detail how these concepts
enter the framework of the isotropic diffusion source approximation,
which we design for the transport of massless fermions through a compressible
gas. Its connection to the well-known diffusion limit is made in
Appendix A. In Sect. \ref{sec:verification}, we evaluate the performance of
this approximation in comparison with Boltzmann neutrino transport
in spherical symmetry. Finally, in Sect. \ref{sec:multid}, we discuss
the extension to multi-dimensional simulations. Details of the finite
differencing and implementation are given in Appendix B.

\section{The isotropic diffusion source approximation (IDSA)}

\label{sec:idsa}In the IDSA,
the separation into hydrodynamics and radiative transfer is not based
on particle species, but on the local opacity. One particle species
is allowed to have a component that evolves in the hydrodynamic limit,
while another component of the same particle species is treated by
radiative transfer. The restriction of a chosen radiative transfer
algorithm to the more transparent regimes enables the use of more
efficient techniques that would not be stable in the full domain.
In opaque regimes, on the other hand, one can take advantage of equilibrium
conditions to reduce the number of primitive variables that need to
be evolved. This algorithmic flexibility can drastically decrease
the overall computational cost with respect to a traditional approach.

In the IDSA, we decompose the
distribution function of one particle species, \( f \), into an isotropic
distribution function of trapped particles, \( f^{{\rm t}} \), and a distribution
function of streaming particles, \( f^{{\rm s}} \). In terms of a linear
operator \( D\left( \right)  \) describing particle propagation,
the transport equation is written as \( D\left( f=f^{{\rm t}}+f^{{\rm s}}\right) =C \),
where \( C=C^{{\rm t}}+C^{{\rm s}} \) is a suitable decomposition of the collision
integral according to the coupling to the trapped (\( C^{{\rm t}} \)) or
streaming (\( C^{{\rm s}} \)) particle components. The ansatz\begin{eqnarray}
D\left( f^{{\rm t}}\right)  & = & C^{{\rm t}}-\Sigma \label{eq:schematic.trapped} \\
D\left( f^{{\rm s}}\right)  & = & C^{{\rm s}}+\Sigma ,\label{eq:schematic.streaming} 
\end{eqnarray}
 requires that we specify an additional source term \( \Sigma  \),
which converts trapped particles into streaming particles and vice
versa. We determine it approximately from the requirement that the
temporal change of \( f^{{\rm t}} \) in Eq. (\ref{eq:schematic.trapped})
has to reproduce the diffusion limit in the limit of small mean free
paths. Hence, we call \( \Sigma  \) the 'diffusion source'. In regions
of large mean free paths, we limit the diffusion source by the local
particle emissivity. Once \( \Sigma  \) is determined by the solution
of Eq. (\ref{eq:schematic.trapped}) for the trapped particle component,
we calculate the streaming particle flux according to Eq. (\ref{eq:schematic.streaming})
by integrating its source, \( C^{{\rm s}}+\Sigma  \), over space. Finally,
the streaming particle distribution function \( f^{{\rm s}} \) is determined
from the quotient of the net particle flux and a geometric estimate
of the flux factor. The diffusion source will turn out to have an
additional weak dependence on \( f^{{\rm s}} \). Thus, iterations or information
from past time steps will be used in the above sequence to reach a
consistent solution.

\subsection{Application to radiative transfer of massless particles}

As our target application is neutrino transport in core collapse supernovae,
we develop and test the IDSA using
the example of the O(\( v/c \)) Boltzmann equation in spherical symmetry
\citep{Lindquist:1966,Castor:1972,Mezzacappa.Bruenn:1993a},
\begin{eqnarray}
\frac{df}{cdt} & + & \mu \frac{\partial f}{\partial r} + \left[ \mu \left( \frac{d\ln \rho }{cdt}+\frac{3v}{cr}\right) +\frac{1}{r}\right] \left( 1-\mu ^{2}\right) \frac{\partial f}{\partial \mu }\nonumber \\
 & + & \left[ \mu ^{2}\left( \frac{d\ln \rho }{cdt}+\frac{3v}{cr}\right) -\frac{v}{cr}\right] E\frac{\partial f}{\partial E}\nonumber \\
 & = & j\left( 1-f\right) -\chi f + \frac{E^{2}}{c\left( hc\right) ^{3}} \nonumber \\
& \times & \left[ \left( 1-f\right) \int Rf'd\mu '-f\int R\left( 1-f'\right) d\mu '\right] .\label{eq:boltzmann} 
\end{eqnarray}
This transport equation describes the propagation of massless fermions at the speed of
light, \( c \), with respect to a compressible background matter
having a rest mass density \( \rho  \). The particle distribution
function \( f\left( t,r,\mu ,E\right)  \) depends on the time, \( t \),
radius, \( r \), and the momentum phase space spanned by the angle
cosine, \( \mu  \), of the particle propagation direction with respect
to the radius and the particle energy, \( E \). The momentum phase
space variables are measured in the frame comoving with the background
matter, which moves with velocity \( v \) with respect to the laboratory
frame. We denote the Lagrangian time derivative in the comoving frame
by \( df/dt \). Note that the derivatives \( \partial f/\partial \mu  \)
and \( \partial f/\partial E \) in Eq. (\ref{eq:boltzmann}) are
also understood to be taken comoving with a fluid element. The particle
density is given by an integration of the distribution function over
the momentum phase space, \( n\left( t,r\right) =4\pi /\left( hc\right) ^{3}\int f\left( t,r,\mu ,E\right) E^{2}dEd\mu  \),
where \( h \) denotes Plancks constant. On the right hand side, we
include a particle emissivity, \( j \), and a particle absorptivity,
\( \chi  \), as well as an isoenergetic scattering kernel, \( R \).
We write out all blocking factors \( \left( 1-f\right)  \) in Eq.
(\ref{eq:boltzmann}) to ease the identification of in-scattering
and out-scattering terms. The shorthand notation \( f' \) refers
to \( f\left( t,r,\mu ',E\right)  \), where \( \mu ' \) is the angle
cosine over which the integration is performed. For the present state
of our approximation we neglect inelastic scattering.

\subsection{Trapped particles}

We separate the particles described by the distribution function \( f=f^{{\rm t}}+f^{{\rm s}} \)
in Eq. (\ref{eq:boltzmann}) into a 'trapped particle' component,
described by a distribution function \( f^{{\rm t}} \), and a 'streaming
particle' component, described by a distribution function \( f^{{\rm s}} \).
We assume that the two components evolve separately according to
Eq. (\ref{eq:boltzmann}), coupled only by an as yet unspecified source
function \( \Sigma  \) which converts trapped particles into streaming
ones and vice versa. In this subsection we discuss the evolution equation
of the trapped particle component,

\begin{eqnarray}
\frac{df^{{\rm t}}}{cdt}+\mu \frac{\partial f^{{\rm t}}}{\partial r} & + & \left[ \mu \left( \frac{d\ln \rho }{cdt}+\frac{3v}{cr}\right) +\frac{1}{r}\right] \left( 1-\mu ^{2}\right) \frac{\partial f^{{\rm t}}}{\partial \mu }\nonumber \\
 & + & \left[ \mu ^{2}\left( \frac{d\ln \rho }{cdt}+\frac{3v}{cr}\right) -\frac{v}{cr}\right] E\frac{\partial f^{{\rm t}}}{\partial E}\nonumber \\
 & = & j-\left( j+\chi \right) f^{{\rm t}}-\Sigma \nonumber \\
 & + & \frac{E^{2}}{c\left( hc\right) ^{3}}\left[ \int Rf^{{\rm t}\prime }d\mu '-f^{{\rm t}}\int Rd\mu '\right] .\label{eq:boltzmann.trapped} 
\end{eqnarray}

\begin{figure}
{\centering \resizebox*{\columnfigurewidth}{!}{\includegraphics{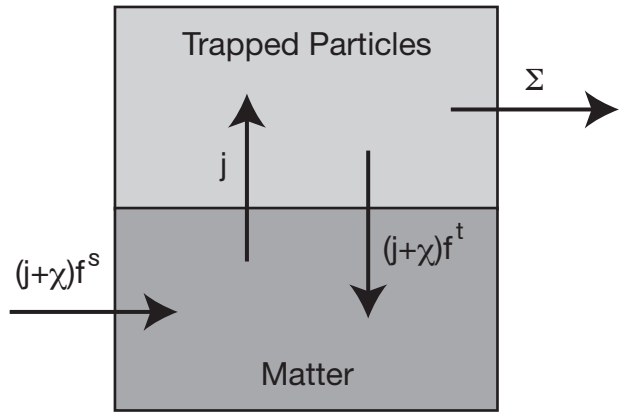}} \par}

\caption{The shaded box schematically represents a fluid element in the diffusion
source approximation. It contains matter (lower part) and trapped
radiation particles (upper part). The interaction with other fluid
elements can exclusively occur by the exchange of streaming particles
or the combined hydrodynamics of the matter and trapped particles.
Hence, streaming particles can be absorbed in matter at the rate \protect\( \left( j+\chi \right) f^{{\rm s}}\protect \)
and trapped particles are converted to streaming particles at the
rate \protect\( \Sigma \protect \). Within the fluid element, matter
emits trapped particles at the rate \protect\( j\protect \) and absorbs
trapped particles at the rate \protect\( \left( j+\chi \right) f^{{\rm t}}\protect \).
The emissivity in the absorption term originates from the identity
\protect\( j\left( 1-f\right) -\chi f=j-\left( j+\chi \right) f\protect \),
which hides the Pauli blocking factor in the absorption term.\label{fig:fluid.element} }
\end{figure}

We assume that the distribution of the trapped particle
component, \( f^{{\rm t}}=f^{{\rm t}}\left( t,r,E\right)  \), and the source
function, \( \Sigma  \), are isotropic. The angular integration of
Eq. (\ref{eq:boltzmann.trapped}) then reduces to\begin{equation}
\label{eq:trapped.isotropic}
\frac{df^{{\rm t}}}{cdt}+\frac{1}{3}\frac{d\ln \rho }{cdt}E\frac{\partial f^{{\rm t}}}{\partial E}=j-\left( j+\chi \right) f^{{\rm t}}-\Sigma .
\end{equation}
 However, even if we are now steering towards the hydrodynamic limit,
the evolution of the trapped particle distribution function in our
approximation should at least reproduce the correct diffusion limit.
The physical understanding of diffusion relies on fast-moving particles
with a very short transport mean free path (see the derivation of
the diffusion limit, Eq. (\ref{eq:diffusion}), in Appendix A). The divergence of the small
net particle flux leads to a slow drain (or replenishment) of particles
in a fluid element. In order to accomodate this diffusive drain (or
replenishment) of trapped particles in Eq. (\ref{eq:trapped.isotropic}),
we have to implement it through the so far unspecified source term \( \Sigma  \).
In the framework of our approximation, trapped particles are converted
to streaming ones (or vice versa) at the same rate diffusion would
drain (or replenish) particles in the fluid element. A comparison
of Eq. (\ref{eq:trapped.isotropic}) with Eq. (\ref{eq:diffusion}) in Appendix A
suggests the following diffusion source,\begin{equation}
\label{eq:definition.sigma.1}
\Sigma =\frac{1}{r^{2}}\frac{\partial }{\partial r}\left( \frac{-r^{2}}{3\left( j+\chi +\phi \right) }\frac{\partial f^{{\rm t}}}{\partial r}\right) +\left( j+\chi \right) \frac{1}{2}\int f^{{\rm s}}d\mu .
\end{equation}
Isoenergetic scattering enters Eq. (\ref{eq:trapped.isotropic}) only
via its opacity \( \phi \) in the expression for the transport mean free path
\(\lambda =  1/\left( j+\chi + \phi \right)\) that determines the flux in Eq. (\ref{eq:definition.sigma.1})
(see Appendix A). The additional term \( \left( j+\chi \right) /2\int f^{{\rm s}}d\mu  \)
accounts for the absorption of streaming particles in matter. Its
necessity is best understood with the help of Fig. (\ref{fig:fluid.element}).
Shown are the different particle fluxes that relate to one fluid element.
The fluid element contains both matter and trapped particles. They
interact by the emissivity \( j \) and absorption \( \left( j+\chi \right) f^{{\rm t}} \)
(vertical arrows). The radiation particles leaving the fluid element
will convert into streaming particles at the rate \( \Sigma  \) (upper
horizontal arrow). We do not include the possibility of a direct emission
of matter into the streaming particle component because this process
is more easily accounted for by a high conversion rate \( \Sigma  \)
of trapped particles. On the other hand, we account for streaming
particles that are absorbed by matter as represented by the arrow
denoted by \( \left( j+\chi \right) f^{{\rm s}} \) in Fig. (\ref{fig:fluid.element}).
In the diffusion limit, the net particle exchange of the fluid element
with its environment, \( \Sigma -\left( j+\chi \right) /2\int f^{{\rm s}}d\mu  \),
must correspond to the diffusion term in Eq. (\ref{eq:diffusion}).
This is the case for our choice of \( \Sigma  \) in Eq. (\ref{eq:definition.sigma.1}).

If the above scheme is applied in a more transparent regime where the
diffusion approximation does not hold, the diffusion source in Eq.
(\ref{eq:definition.sigma.1}) may become arbitrarily large. This
would be inconsistent with the particle fluxes drawn in Fig. (\ref{fig:fluid.element})
because over a long time the diffusion source can not exceed the emissivity
of trapped particles without creating an unphysical deficit in trapped
particles. Instead, we would like the trapped particle component to
drop to zero and stay zero in the limit of long mean free paths. This
is achieved if we limit the diffusion source in Eq. (\ref{eq:definition.sigma.1})
to \( \Sigma \leq j \). If the diffusion source and emissivity reach
equality, the matter absorptivity -\( \left( j+\chi \right) f^{{\rm t}} \)
removes remaining trapped particles while all newly emitted ones are
directly converted to streaming particles that escape the fluid element.
With this limit imposed, the net interaction of particles with matter
in Fig. (\ref{fig:fluid.element}) has the correct limit for large
mean free paths, \( j-\left( j+\chi \right) f^{{\rm s}} \).

The search for a lower bound to the diffusion source is less straight-forward.
In principle, a negative \( \Sigma  \) can not be physically excluded.
It corresponds to streaming particles that become trapped in a region
of large opacity and low absorptivity. A limit corresponding to the
restriction \( f^{{\rm t}}\leq 1 \) would suggest \( \Sigma \geq -\chi  \).
However,
\( f^{{\rm t}}=1 \) can be an excessively large particle density compared
to the physically expected value. Much more stable and more accurate
results were obtained by the requirement that \( f^{{\rm t}}\leq j/\left( j+\chi \right)  \),
where \( j/\left( j+\chi \right)  \) represents the equilibrium distribution
function. If one considers this condition for the particle fluxes
in Fig. (\ref{fig:fluid.element}) this leads to the simple requirement
\( \Sigma \geq 0 \). Hence, the net absorption of particles in a
fluid element can not exceed \( \left( j+\chi \right) f^{{\rm s}} \). The
diffusion source from Eq. (\ref{eq:definition.sigma.1}) then becomes\begin{eqnarray}
\Sigma  & = & \min \left\{ \max \left[ \alpha +\left( j+\chi \right) \frac{1}{2}\int f^{{\rm s}}d\mu ,0\right] ,j\right\} \nonumber \\
\alpha  & = & \frac{1}{r^{2}}\frac{\partial }{\partial r}\left( \frac{-r^{2}}{3\left( j+\chi +\phi \right) }\frac{\partial f^{{\rm t}}}{\partial r}\right) .\label{eq:definition.sigma.2} 
\end{eqnarray}
Assuming that the streaming particle density \( 1/2\int f^{{\rm s}}d\mu  \)
was known from the considerations described in the following
subsection, Eq. (\ref{eq:trapped.isotropic}) can be used to calculate
the evolution of the trapped particle component consistently with
the diffusion source specified in Eq. (\ref{eq:definition.sigma.2}).

\subsection{Streaming particles}

The evolution equation for the streaming particle component consists
of all terms in Eq. (\ref{eq:boltzmann}) that have not been considered
in the evolution equation (\ref{eq:boltzmann.trapped}) for the trapped
particles. As mentioned above, we neglect direct scattering from the trapped
component into the streaming component and vice versa. That is
\begin{eqnarray}
\frac{df^{{\rm s}}}{cdt} & + & \mu \frac{\partial f^{{\rm s}}}{\partial r} + \left[ \mu \left( \frac{d\ln \rho }{cdt}+\frac{3v}{cr}\right) +\frac{1}{r}\right] \left( 1-\mu ^{2}\right) \frac{\partial f^{{\rm s}}}{\partial \mu }\nonumber \\
& + & \left[ \mu ^{2}\left( \frac{d\ln \rho }{cdt}+\frac{3v}{cr}\right) -\frac{v}{cr}\right] E\frac{\partial f^{{\rm s}}}{\partial E} = -\left( j+\chi \right) f^{{\rm s}} \nonumber \\
& + & \Sigma +\frac{E^{2}}{c\left( hc\right) ^{3}}\left[ \int Rf^{{\rm s}\prime }d\mu '-f^{{\rm s}}\int Rd\mu '\right] .\label{eq:boltzmann.streaming} 
\end{eqnarray}
For the evolution of streaming particles we neglect the scattering
integrals on the right hand side of Eq. (\ref{eq:boltzmann.streaming}),
because the streaming particle density is designed to be small compared
to the trapped particle density in regions where the scattering rate
dominates the interactions. The weak coupling between streaming particles
and the background matter in more transparent regimes where the streaming
particle component becomes large makes an inertial laboratory frame
more convenient for the solution of Eq. (\ref{eq:boltzmann.streaming})
than the frame comoving with the fluid. One may simply substitute
the conditions \( d\ln \rho /cdt=0 \) and \( v=0 \) for a static
background in Eq. (\ref{eq:boltzmann.streaming}) to find the streaming
particle evolution equation in the laboratory frame:\begin{eqnarray}
\frac{\partial \hat{f}^{{\rm s}}}{c\partial \hat{t}}+\hat{\mu }\frac{\partial \hat{f}^{{\rm s}}}{\partial r} & + & \frac{1}{r}\left( 1-\hat{\mu }^{2}\right) \frac{\partial \hat{f}^{{\rm s}}}{\partial \hat{\mu }}=-\left( \hat{j}+\hat{\chi }\right) \hat{f}^{{\rm s}}+\hat{\Sigma }.\label{eq:free.streaming} 
\end{eqnarray}
Note that the quantities carrying a hat now are measured
in the laboratory frame. In this frame, the particle energy is a constant
of motion. The cumbersome energy derivatives of the distribution function
in Eq. (\ref{eq:boltzmann.streaming}) are avoided, but some quantities
require Lorentz transformations between the comoving and laboratory
frames. 

In contrast to the full Boltzmann equation (\ref{eq:boltzmann}),
the source term on the right hand side couples only weakly to the
particle distribution function \( \hat{f}^{{\rm s}} \). In the diffusive
domain, the angle integrated term \( -\left( \hat{j}+\hat{\chi }\right) \hat{f}^{{\rm s}} \)
cancels to a large extent with the corresponding contribution to \( \hat{\Sigma } \)
as defined in Eq. (\ref{eq:definition.sigma.2}). In the free streaming
domain, the emissivity \( \hat{j} \) and opacity \( \hat{\chi } \) are assumed
to be small.
Depending on the application, there might be various suitable approaches
to solve Eq. (\ref{eq:free.streaming}). Here,
we will proceed with a stationary-state approximation. If we assume
that the fluxes of free streaming particles reach stationary values
much faster than the dynamical time scale of interest, we can drop
the time-derivative in the first term in Eq. (\ref{eq:free.streaming}).
If the source on the right hand side is assumed to be known from a
consistent solution of Eqs. (\ref{eq:trapped.isotropic}) and (\ref{eq:definition.sigma.2}),
the integration of Eq. (\ref{eq:free.streaming}) over angles leads
to a Poisson equation for a potential \( \psi  \), whose gradient
represents the particle flux: \begin{eqnarray}
\frac{\partial \psi }{\partial r} & = & \frac{1}{2}\int \hat{f}^{{\rm s}}\hat{\mu }d\hat{\mu }\nonumber \\
\frac{1}{r^{2}}\frac{\partial }{\partial r}\left( r^{2}\frac{\partial \psi }{\partial r}\right)  & = & \frac{1}{2}\int \left[ -\left( \hat{j}+\hat{\chi }\right) \hat{f}^{{\rm s}}+\hat{\Sigma }\right] d\hat{\mu }.\label{eq:spherical.poisson} 
\end{eqnarray}

In the end, we are interested in the streaming particle density rather
than flux. Both quantities are related by the flux factor. It is a
very challenging problem to calculate an accurate flux factor. However,
Bruenn in \citep{Liebendoerfer.Messer.ea:2004} suggested a simple
and very successful approximation based on the assumption that all
particles of a given energy are isotropically emitted at their corresponding
scattering sphere. This leads to the additional relation
\begin{equation}
\label{eq:flux.factor}
\frac{1}{2}\int \hat{f}^{{\rm s}}\left( E\right)d\hat{\mu }=
\frac{2\frac{\partial \psi }{\partial r}\left( E\right)}
{1+\sqrt{1-\left(
\frac{R_{\nu }\left( E\right)}
{\max \left( r,R_{\nu }\left( E\right)\right)} \right) ^{2}}},
\end{equation}
where \( R_{\nu }\left( E\right) \) is the radius of the monochromatic scattering sphere
that depends on the particle energy \( E\).

In order to solve for both trapped and streaming particle components,
we have to Lorentz-transform quantities between the comoving and laboratory
frames. The source of Eq. (\ref{eq:spherical.poisson}) in the laboratory
frame is related to the comoving frame by
\begin{eqnarray}
& & \frac{1}{2}\int \left[ -\left( \hat{j}+\hat{\chi }\right) \hat{f}^{{\rm s}}+\hat{\Sigma }\right] d\hat{\mu }\hat{E}^{2}d\hat{E} \nonumber \\
& = & \frac{1}{2}\int \left[ -\left( j+\chi \right) f^{{\rm s}}+\Sigma \right] d\mu E^{2}dE.\label{eq:lorentz.source}
\end{eqnarray}
Here we used the fact that the source is isotropic in the comoving
frame together with the relations \( d\hat{\mu }\hat{E}d\hat{E}=d\mu EdE \),
\( \hat{E}=\gamma \left( 1+\mu v/c\right) E \) and \( \hat{j}\hat{E}=jE \)
\citep{Mihalas.Mihalas:1984}, where \( \gamma =1/\sqrt{1-\left( v/c\right) ^{2}} \)
is the Lorentz factor. On the other hand, the streaming particle contribution
to Eq. (\ref{eq:definition.sigma.2}) in the comoving frame is related
to the solution of Eq. (\ref{eq:spherical.poisson}) in the laboratory
frame by
\begin{eqnarray}
& & \frac{1}{2}\int f^{{\rm s}}d\mu E^{2}dE = \frac{1}{2}\int \hat{f}^{{\rm s}}\gamma \left( 1-\hat{\mu }v/c\right) d\hat{\mu }\hat{E}^{2}d\hat{E}\nonumber \\
& = & \gamma \left( \frac{1}{2}\int \hat{f}^{{\rm s}} d\hat{\mu }-\frac{v}{c} \frac{\partial \psi }{\partial r}\right) \hat{E}^{2}d\hat{E},\label{eq:lorentz.streaming} 
\end{eqnarray}
where the first term is given in Eq. (\ref{eq:flux.factor}). In \( O\left( v/c\right)  \) we may neglect the factor \( \gamma  \).

Hence, as soon as the evolution of the trapped particle component
according to Eq. (\ref{eq:trapped.isotropic}) has determined the
source in Eq. (\ref{eq:definition.sigma.2}), one can transform it
to the laboratory frame by Eq. (\ref{eq:lorentz.source}). Then one
determines the stationary-state streaming particle flux and density
based on Eqs. (\ref{eq:spherical.poisson}) and (\ref{eq:flux.factor})
in the laboratory frame. The resulting streaming particle density
is transformed back to the comoving frame by Eq. (\ref{eq:lorentz.streaming}).
In this form it can be used for the next iteration or time step in
Eq. (\ref{eq:definition.sigma.2}).

\subsection{Coupling with hydrodynamics}

\label{sec:coupling.with.hydrodynamics}The evolution of the trapped
particle component in Eqs. (\ref{eq:trapped.isotropic}) and (\ref{eq:definition.sigma.2})
and the stationary-state limit of the streaming particle component
in Eqs. (\ref{eq:spherical.poisson}) and (\ref{eq:flux.factor})
must be coupled with the dynamics of the background matter. We assume
that the dynamics of the background matter is well described by the
conservation laws of hydrodynamics,\begin{equation}
\label{eq:conservation.law}
\frac{\partial }{\partial t}U+\frac{\partial }{r^{2}\partial r}\left( r^{2}F\right) =0,
\end{equation}
 where \( U \) is a vector of primitive variables and \( F \) a
vector of fluxes. We rewrite the evolution equation (\ref{eq:trapped.isotropic})
for the trapped particle component with an Eulerian time derivative
and use the continuity equation to substitute the \( d\ln \rho /dt \)
term by the velocity divergence,
\begin{eqnarray}
\frac{\partial f^{{\rm t}}}{c\partial t} & + & \frac{\partial }{r^{2}\partial r}\left( r^{2}\frac{v}{c}f^{{\rm t}}\right) -\frac{\partial }{r^{2}\partial r}\left( r^{2}\frac{v}{c}\right) \frac{\partial \left( E^{3}f^{{\rm t}}\right) }{3E^{2}\partial E}\nonumber \\
& = & j-\left( j+\chi \right) f^{{\rm t}}-\Sigma . \label{eq:trapped.advection}
\end{eqnarray}
 One could now merge the hydrodynamical update on the left
hand side of Eq. (\ref{eq:trapped.advection}) with the conservation
law in Eq. (\ref{eq:conservation.law}). However, if the dynamics
of the background matter requires a large-scale numerical simulation
and if there are many particle species, the required memory and CPU
time to store and advect all these quantities can easily exceed the
capacity of the available hardware. In our supernova application,
for example, we need in three dimensions a vector \( U \) of \( 6 \)
entries to describe the hydrodynamics as detailed below. The neutrino
transport involves \( 2 \)-\( 4 \) neutrino species (\( \nu _{{\rm e}} \),
\( \bar{\nu }_{{\rm e}} \), \( \nu _{\mu /\tau } \), \( \bar{\nu }_{\mu /\tau } \))
with at least \( 12 \) entries per species to sample the different
neutrino energies \( E \) in \( f^{{\rm t}}\left( t,r,E\right)  \).

If the characteristic time scale of local reactions between the transported
particle species is faster than the diffusion time scale, one can
use equilibrium conditions to reduce the number of primitive variables
necessary to describe the distribution functions of trapped particles.
In the supernova model, for example, we can approximate the spectrum
by a thermal spectrum. Note, that this assumption is only made for
the \emph{trapped} particles within a fluid element. The streaming
particles, which communicate \emph{between} fluid elements, keep their
detailed spectral information. Our scheme should therefore not be
confused with a 'grey' scheme that may additionally assume a predefined
energy-spectrum for the particle exchange between fluid elements. We characterise
the thermal equilibrium by a particle number fraction, \( Y^{{\rm t}} \),
and a particle mean specific energy, \( Z^{{\rm t}} \),\begin{eqnarray}
Y^{{\rm t}} & = & \frac{m_{\rm b}}{\rho }\frac{4\pi }{\left( hc\right) ^{3}}\int f^{{\rm t}}E^{2}dEd\mu \nonumber \\
Z^{{\rm t}} & = & \frac{m_{\rm b}}{\rho }\frac{4\pi }{\left( hc\right) ^{3}}\int f^{{\rm t}}E^{3}dEd\mu ,\label{eq:definition.yt.zt} 
\end{eqnarray}
where \( m_{\rm b} \) is a constant relating a particle number to the
rest mass of the background matter (in our application it is the baryon
mass).

The corresponding energy integrals performed on Eq. (\ref{eq:trapped.advection})
lead to evolution equations for \( Y^{{\rm t}} \) and \( Z^{{\rm t}} \),\begin{eqnarray*}
 &  & \frac{\partial }{\partial t}\left( \rho Y^{{\rm t}}\right) +\frac{\partial }{r^{2}\partial r}\left( r^{2}v\rho Y^{{\rm t}}\right) \\
 & = & m_{\rm b}\frac{4\pi c}{\left( hc\right) ^{3}}\int \left[ j-\left( j+\chi \right) f^{{\rm t}}-\frac{1}{2}\int \Sigma d\mu \right] E^{2}dE\\
 &  & \frac{\partial }{\partial t}\left( \rho Z^{{\rm t}}\right) +\frac{\partial }{r^{2}\partial r}\left( r^{2}v\rho Z^{{\rm t}}\right) +\frac{\partial }{r^{2}\partial r}\left( r^{2}v\right) \frac{\rho Z^{{\rm t}}}{3}\\
 & = & m_{\rm b}\frac{4\pi c}{\left( hc\right) ^{3}}\int \left[ j-\left( j+\chi \right) f^{{\rm t}}-\Sigma \right] E^{3}dE.
\end{eqnarray*}
The equation for \( Z^{{\rm t}} \) corresponds to an energy equation with
a \( pdV \) term for the radiation pressure \( \rho Z^{{\rm t}}/3 \) on
the left hand side. Analogously to an entropy equation it can be cast
into a conservative form for the evolution of \( \left( \rho Z^{{\rm t}}\right) ^{3/4} \).
Hence, one can solve the advective part of Eq. (\ref{eq:trapped.advection})
together with the hydrodynamics conservation law (\ref{eq:conservation.law})
based on the following primitive variables:\begin{equation}
\label{eq:primitive.variables}
U=\left( \begin{array}{c}
\rho \\
\rho v\\
\rho \left( e+\frac{1}{2}v^{2}\right) \\
\rho Y_{{\rm e}}\\
\rho Y_{l}^{{\rm t}}\\
\left( \rho Z_{l}^{{\rm t}}\right) ^{\frac{3}{4}}
\end{array}\right) ,\quad F=\left( \begin{array}{c}
v\rho \\
v\rho v+p\\
v\rho \left( e+\frac{1}{2}v^{2}+\frac{p}{\rho }\right) \\
v\rho Y_{{\rm e}}\\
v\rho Y_{l}^{{\rm t}}\\
v\left( \rho Z_{l}^{{\rm t}}\right) ^{\frac{3}{4}}
\end{array}\right) ,
\end{equation}
 where \( p \) is the fluid pressure, \( e \) the fluid specific
internal energy and \( Y_{{\rm e}} \) the electron fraction. The index
\( l \) labels different species of trapped particles. 

As the distribution function of trapped particles in thermal equilibrium,
\( f_{l}^{{\rm t}}\left( E\right) =\left\{ \exp \left[ \beta _{l}\left( E-\mu _{l}\right) \right] +1\right\} ^{-1} \),
has two free parameters \( \beta _{l} \) and \( \mu _{l} \), it
can be reconstructed so that Eq. (\ref{eq:definition.yt.zt}) is fulfilled
for the new values of \( Y_{l}^{{\rm t}} \) and \( Z_{l}^{{\rm t}} \). Then,
the remaining update of the trapped particle distribution in Eq. (\ref{eq:trapped.advection})
is given by \begin{equation}
\label{eq:trapped.update.f}
\frac{\partial f_{l}^{{\rm t}}}{c\partial t}=j_{l}-\left( j_{l}+\chi _{l}\right) f_{l}^{{\rm t}}-\Sigma _{l}.
\end{equation}
This equation also determines the net interaction rates, \( s_{l}=j_{l}-\left( j_{l}+\chi _{l}\right) \left( f_{l}^{{\rm t}}+f_{l}^{{\rm s}}\right)  \),
between matter and the radiation particles, which leads to the following
changes of the electron fraction and internal specific energy:\begin{eqnarray}
s_{l} & = & \frac{\partial f_{l}^{{\rm t}}}{c\partial t}+\Sigma _{l}-\left( j_{l}+\chi _{l}\right) \frac{1}{2}\int f_{l}^{{\rm s}}d\mu \label{eq:trapped.update.matter} \\
\frac{\partial Y_{{\rm e}}}{c\partial t} & = & -\frac{m_{\rm b}}{\rho }\frac{4\pi c}{\left( hc\right) ^{3}}\int \left( s_{\nu _{{\rm e}}}-s_{\bar{\nu }_{{\rm e}}}\right) E^{2}dE\label{eq:trapped.update.ye} \\
\frac{\partial e}{c\partial t} & = & -\frac{m_{\rm b}}{\rho }\frac{4\pi c}{\left( hc\right) ^{3}}\int \left( s_{\nu _{{\rm e}}}+s_{\bar{\nu }_{{\rm e}}}\right) E^{3}dE.\label{eq:trapped.update.e} 
\end{eqnarray}
The changes in the matter electron fraction and specific energy feed
back into the emissivity and absorptivity used in Eq. (\ref{eq:trapped.update.f}).
Once a consistent solution has been found, Eq. (\ref{eq:trapped.update.f})
allows updates of the trapped particle fraction, of the trapped particle
specific energy, and of the matter velocity, which is subject to radiation
pressure. \begin{eqnarray}
\frac{\partial Y_{l}^{{\rm t}}}{\partial t} & = & \frac{m_{\rm b}}{\rho }\frac{4\pi c}{\left( hc\right) ^{3}}\int \frac{\partial f_{l}^{{\rm t}}}{\partial t}E^{2}dE\label{eq:trapped.update.yt} \\
\frac{\partial Z_{l}^{{\rm t}}}{\partial t} & = & \frac{m_{\rm b}}{\rho }\frac{4\pi c}{\left( hc\right) ^{3}}\int \frac{\partial f_{l}^{{\rm t}}}{\partial t}E^{3}dE\label{eq:trapped.update.zt} \\
\frac{\partial v}{\partial t} & = & -\frac{1}{\rho }\frac{\partial }{\partial r}\left( \frac{\rho Z_{l}^{{\rm t}}}{3m_{\rm b}}\right) .\label{eq:trapped.update.velocity} 
\end{eqnarray}
Finally, the cycle of updates is completed by the solution of Eqs.
(\ref{eq:spherical.poisson}) and (\ref{eq:flux.factor}) for the
distribution function \( f^{{\rm s}} \) for streaming particles based on
the sources determined in Eqs. (\ref{eq:trapped.update.f}) and (\ref{eq:definition.sigma.2}).
Details on the finite differencing of our implementation of the IDSA are given in Appendix B.

\section{Verification for spherically symmetric supernova models}

\label{sec:verification}We implement the IDSA in the hydrodynamics code Agile
\citep{Liebendoerfer.Rosswog.Thielemann:2002}. For the evaluation
of the radiative transfer approximations, we run the simulations in the
Newtonian limit and include only electron
flavour neutrinos. Their dominant emission and absorption reactions
are\begin{eqnarray}
e^{-}+p & \rightleftharpoons  & n+\nu _{{\rm e}} \nonumber\\
e^{+}+n & \rightleftharpoons  & p+\bar{\nu }_{{\rm e}},
\label{eq:reactions}
\end{eqnarray}
where \( e^{-} \), \( e^{+} \), \( \nu _{{\rm e}} \), \( \bar{\nu }_{{\rm e}} \)
refer to electrons, positrons, electron neutrinos and electron antineutrinos,
respectively, that interact with protons, \( p \), and neutrons \( n \).
The opacities in our example are given by isoenergetic scattering
on nucleons and nuclei. All weak interactions are implemented as described
in \citep{Bruenn:1985}. The thermodynamical state of matter as a function
of density, \( \rho  \), temperature, \( T \), and electron fraction,
\( Y_{{\rm e}} \), is calculated by the Lattimer-Swesty equation of state
\citep{Lattimer.Swesty:1991}.

Very simple and efficient approximations to treat the deleptonisation
in the collapse phase have been suggested and evaluated in an earlier
paper \citep{Liebendoerfer:2005}.
Here we concentrate on the postbounce phase, which is more difficult
to capture with neutrino physics approximations. The neutrinos not only
leak out, but they also interact at distant locations and feed back
into the hydrodynamics by their transfer of lepton number and energy.
Due to the as yet restricted input physics, we base our first comparison
on model N13 in \citep{Liebendoerfer.Rampp.ea:2005}, which we have
repeated with the most recent code version to obtain perfect consistency
in all parts that are not related to the neutrino transport. The additional
neutrino-electron scattering and pair creation rates, which are included
in model N13, do not significantly contribute to the electron flavour
neutrino transport in the postbounce phase. In order to keep the approximation
as simple as possible, we additionally neglect the transformations between
comoving and laboratory frames in our current implementation.

\begin{figure}
{\centering \resizebox*{\columnfigurewidth}{!}{\includegraphics{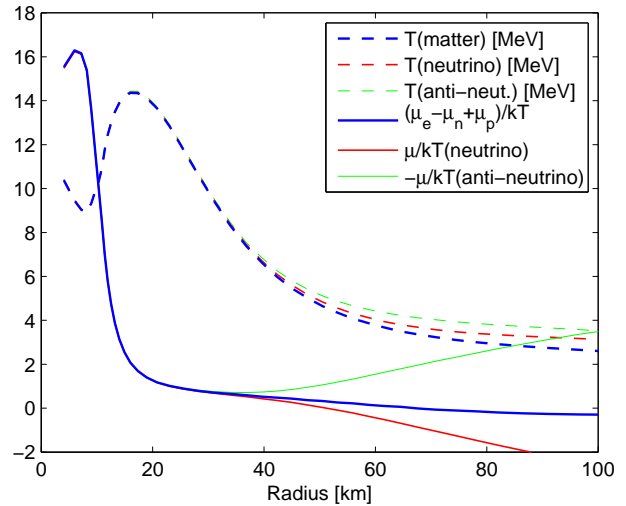}} \par}

\caption{Neutrino temperature and degeneracy parameters resulting
from the application of Eq. (\protect\ref{eq:definition.yt.zt}) at a typical time of
\protect\( 150 \protect\) ms after bounce. As indicated in the legend, the solid
lines show the degeneracy of the particles involved in Eq.
(\protect\ref{eq:reactions}) together with the (anti-)\-neutrino degeneracy parameters.
The dashed lines show the matter temperature together with the
(anti-)neutrino temperature parameters.}
\label{fig:fig2.ps}
\end{figure}
We start with the verification of the description of the trapped particle component.
The trapped particle distribution function is assumed to be thermal and is parameterised
by the two parameters inverse temperature, \( \beta \), and the particle degeneracy
parameter, \( \mu /(kT) \). They are chosen such that Eq. (\ref{eq:definition.yt.zt}) is
fulfilled. In Fig. \ref{fig:fig2.ps} we compare these parameters to the temperature of matter
and to the equilibrium degeneracy parameter determined by reactions (\ref{eq:reactions}).
We find as expected that the neutrino distribution function of a reference simulation
with Boltzmann neutrino transport is very well described by the parameterisation in
regimes where the neutrinos are trapped and thermal and weak equilibrium are good
approximations. Deviations are visible where the neutrinos start to decouple from
matter. The IDSA leads to decreasing trapped particle abundances in this transition
regime as more and more particles change from the trapped particle component to
the streaming particle component. This is visible in the decreasing chemical potentials at
radii larger than \( 40 \) km. The temperature parameter of the remaining trapped
particle component is slightly larger than the temperature of the matter.

\begin{figure*}
{\centering \resizebox*{\twocolumnfigurewidth}{!}{\includegraphics{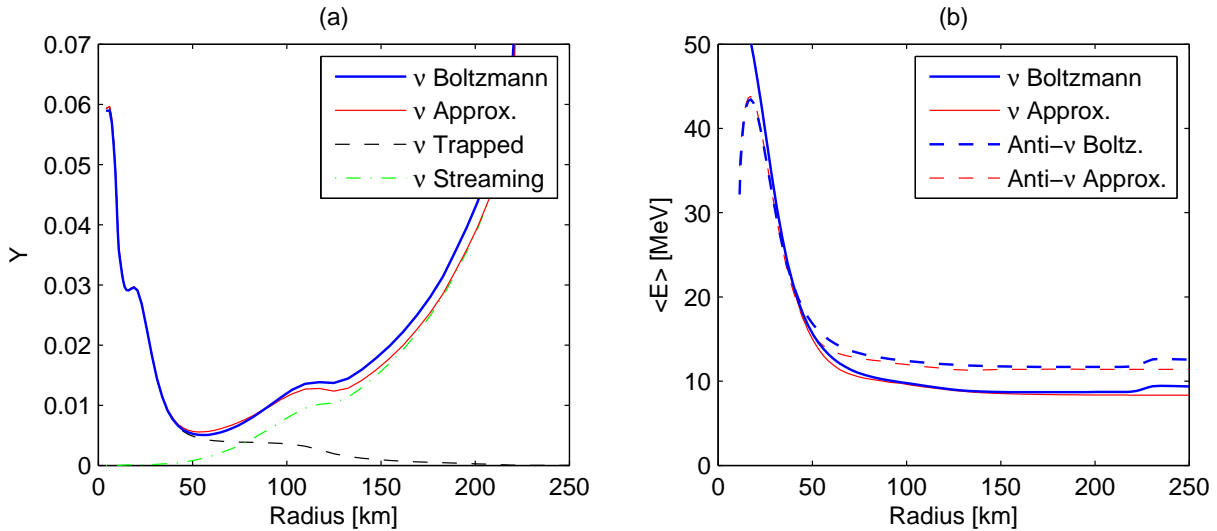}} \par}
\caption{Panel (a) shows neutrino abundances as a function of radius at a typical
postbounce time of \protect\( 150 \) ms. The thick solid line shows the neutrino
abundance in the reference simulation with Boltzmann transport. The
application of the IDSA leads to a trapped neutrino abundance (dashed line) and
a streaming neutrino abundance (dash-dotted line). They sum up to the total
neutrino abundance (thin solid line), which is comparable to
the Boltzmann result. Graph (b) demonstrates that also the neutrino mean energies
between the reference model (thick lines) and the IDSA (thin lines) agree nicely.}
\label{fig:fig3.ps}
\end{figure*}
This decoupling of particles from opaque matter is an important feature that needs to
be handled by a radiative transfer algorithm. Since the IDSA treats this transition
by a gradual conversion of 'trapped particles' to 'streaming particles', we can
compare the sum of the trapped particle and streaming particle densities
with the total particle density in a reference simulation that solves the Boltzmann
equation. Fig. \ref{fig:fig3.ps} shows this comparison for a typical time slice at
 \( 150 \) ms after bounce. Because the trapped particle distribution function in
 Eq. (\ref{eq:definition.yt.zt}) is characterised by the particle abundance and mean
 energy, we select these two quantities for the comparison. Panel (a) shows how
 the trapped and streaming particle abundances
overlap and add up to a reasonable total particle abundance. Panel (b) shows
similar agreement in the particle mean energies. Although the results agree very
well in general, we also note that the deviations are largest around \( 50 \) km radius.
Fig. \ref{fig:fig2.ps} indicates that this is exactly in the difficult regime where the
neutrinos decouple from matter. The less relevant deviations at radii larger than \( 220 \)
km stem from the Doppler energy shift at the shock front, that is not present
in the IDSA data, because the transformation between the laboratory and comoving
frames has been neglected.

\begin{figure}
{\centering \resizebox*{\columnfigurewidth}{!}{\includegraphics{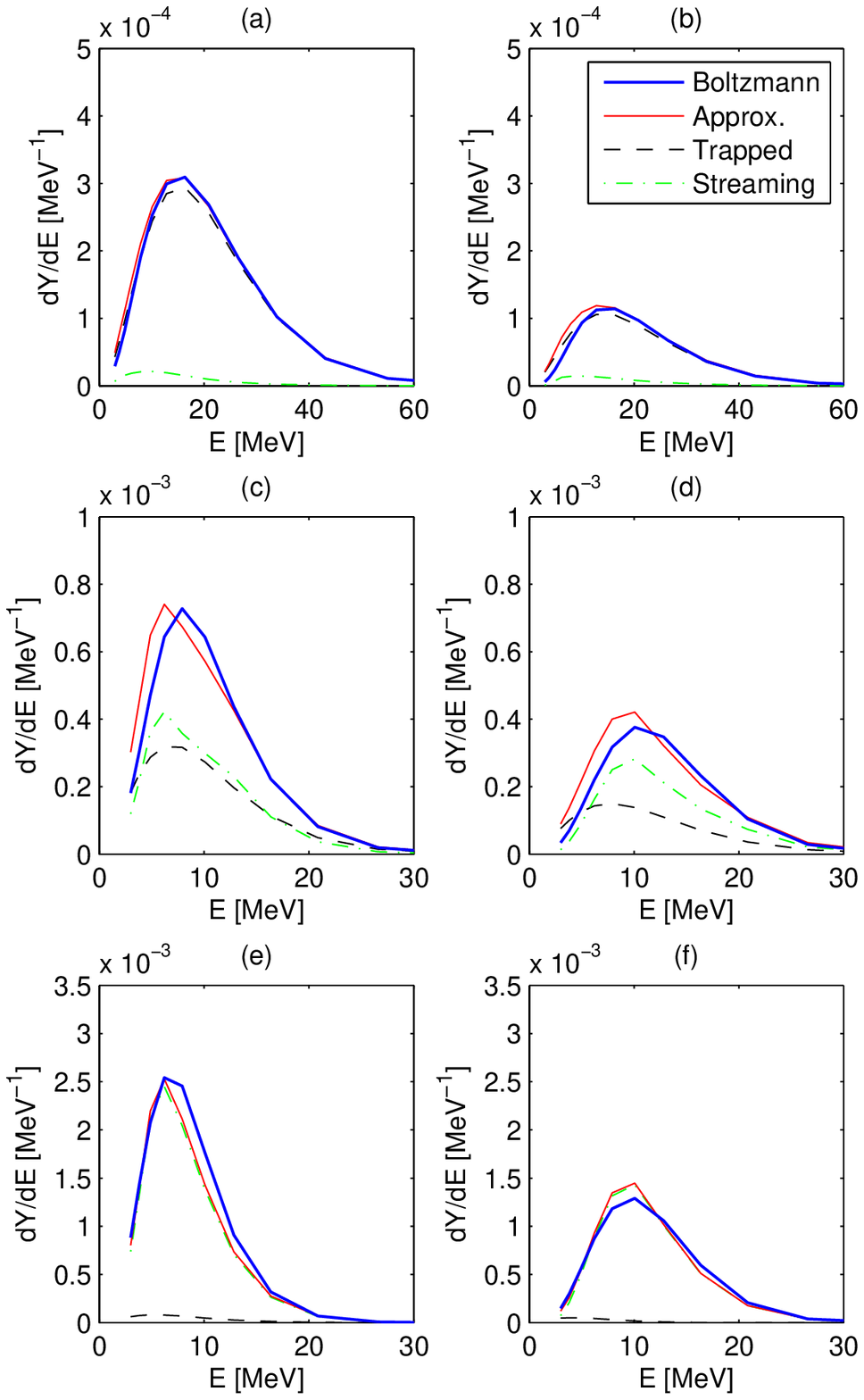}} \par}
\caption{Particle spectra for the trapped particle component (dashed line), the
streaming particle component (dash-dotted line), their sum (thin solid line) and
the reference simulation (thick solid line) at selected radii. Panels (a) and (b) show
spectra at \protect\( 40 \protect\) km radius, panels (c) and (d) at \protect\( 80 \protect\)
km radius and panels (e) and (f) at \protect\( 160 \) km radius. The panels on the left
hand side (a,c,e) show neutrino spectra while the panels on the right hand side
(b,d,f) show antineutrino spectra.}
\label{fig:fig4.ps}
\end{figure}
We can analyse the discrepancies further by looking at the particle spectra in the
different regimes. The panels on the left hand side of Fig. \ref{fig:fig4.ps} show
neutrino spectra, while the panels on the right hand side show antineutrino spectra.
The spectra in panels (a) and (b) are taken at \( 40 \) km radius from the time slice
at \( 150 \) ms after bounce. One can immediately see that the trapped neutrino
component (dashed lines) dominates over the streaming particle component
(dash-dotted lines). The sum of both (thin solid line) matches well with the reference
solution (thick solid line). At these conditions the spectrum of the neutrinos is very
accurately represented by the IDSA, while small differences become visible in the
antineutrino spectra. The spectra match well, because the thermal distribution
function used for the trapped neutrino component is still a good approximation
at the border of the neutrino-opaque regime. Panels (c) and (d) show the same
quantities further out, where the trapped and streaming particle abundances
reach about the same value (i.e. at \( \sim 80 \) km radius). The particle abundances
again match well with the reference data, but the neutrino spectra in the IDSA
peak at lower energies than in the reference simulation. This is consistent with
the findings above in Fig. \ref{fig:fig3.ps}. The reason is that 
the assumption of a thermal spectrum for the trapped particle component starts
to become inaccurate at \( 80 \) km radius, while the trapped particle
component still makes a significant
contribution of \(\sim 50 \)\% to the total abundance. The inaccuracy of a thermal
spectrum in the semi-transparent regime is well-documented in the supernova
literature (e.g. \citet{Myra.Burrows:1990}) and appears most dramatic if presented
on logarithmic axes. We prefer a linear axis where the total particle abundance, or
deviations between abundances, are proportional to the enclosed area between
the lines. At a radius of \( 160 \) km the streaming particle component dominates
over the trapped particle component as shown in panels (e) and (f). The
comparison of the spectra shows more
accurate agreement because the streaming particle component keeps its full
spectral information in the IDSA. Some deviations in the abundances are visible.
They are most likely an effect of the approximations in the decoupling regime
and demonstrate the limitations of the IDSA.

\begin{figure}
{\centering \resizebox*{\columnfigurewidth}{!}{\includegraphics{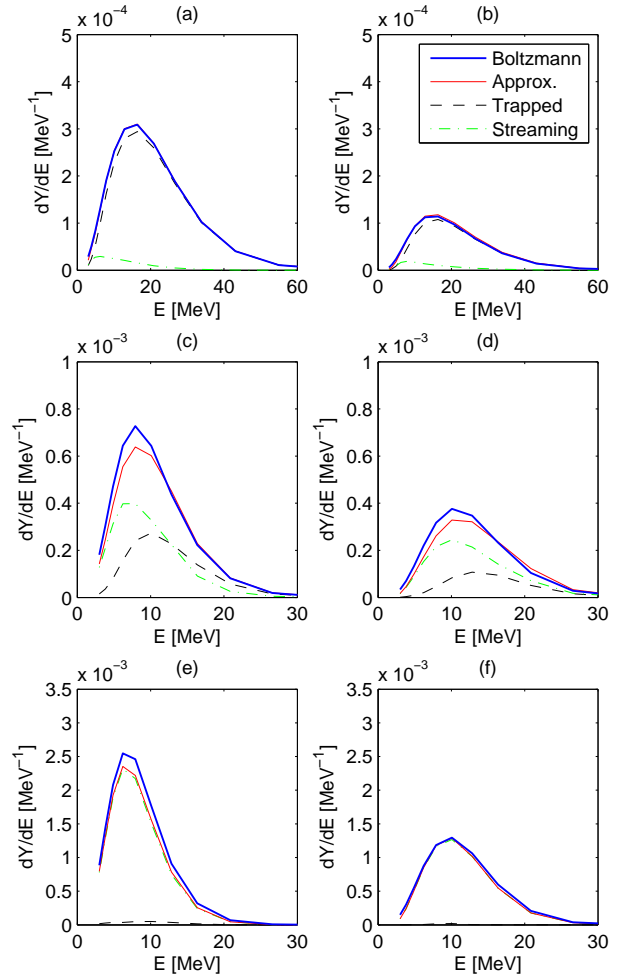}} \par}
\caption{The approximate data in this figure has been calculated with a version of the IDSA
where the spectral information about the trapped particle component is retained as
described in Appendix B. Otherwise, the representation is the same as in Fig.
 \protect\ref{fig:fig4.ps}: Panels (a) and (b) show
spectra at \protect\( 40 \protect\) km radius, panels (c) and (d) at \protect\( 80 \protect\)
km radius and panels (e) and (f) at \protect\( 160 \) km radius. The panels on the left
hand side (a,c,e) show neutrino spectra while the panels on the right hand side
(b,d,f) show antineutrino spectra.}
\label{fig:fig5.ps}
\end{figure}
At this stage we are curious, how much accuracy was lost by the decision to
parameterise the trapped particle component by a thermal distribution function.
In order to test this, we developed an alternative version that takes deviations from
the thermal trapped particle spectrum into account. The details of this test approach
are described in Eq. (\ref{eq:spectral.correction}) in Appendix B.
The resulting spectra of this more complicated version are shown
in Fig.  \ref{fig:fig5.ps}. Now the reference spectra are perfectly matched by the
IDSA results at \( 40 \) km radius and for the antineutrinos also at \( 160 \) km radius.
Even in the difficult regime at \( 80 \) km radius the spectral shapes agree very
nicely with the reference. But new deviations appear this time in the particle abundances,
which are somewhat smaller in the IDSA data than in the reference data. We conclude
from these comparisons that the results of the IDSA are in qualitatively robust
agreement with the reference data, but inaccuracies of \( 10-20 \)\% are likely
to be present in the details of the distribution function and are difficult to
remove by simple upgrades. In order to become able to better distinguish the
systematic errors from the accidental errors, we include both the normal and
the spectral version of the IDSA in the following comparisons with the reference
data.

\begin{figure}
{\centering \resizebox*{\columnfigurewidth}{!}{\includegraphics{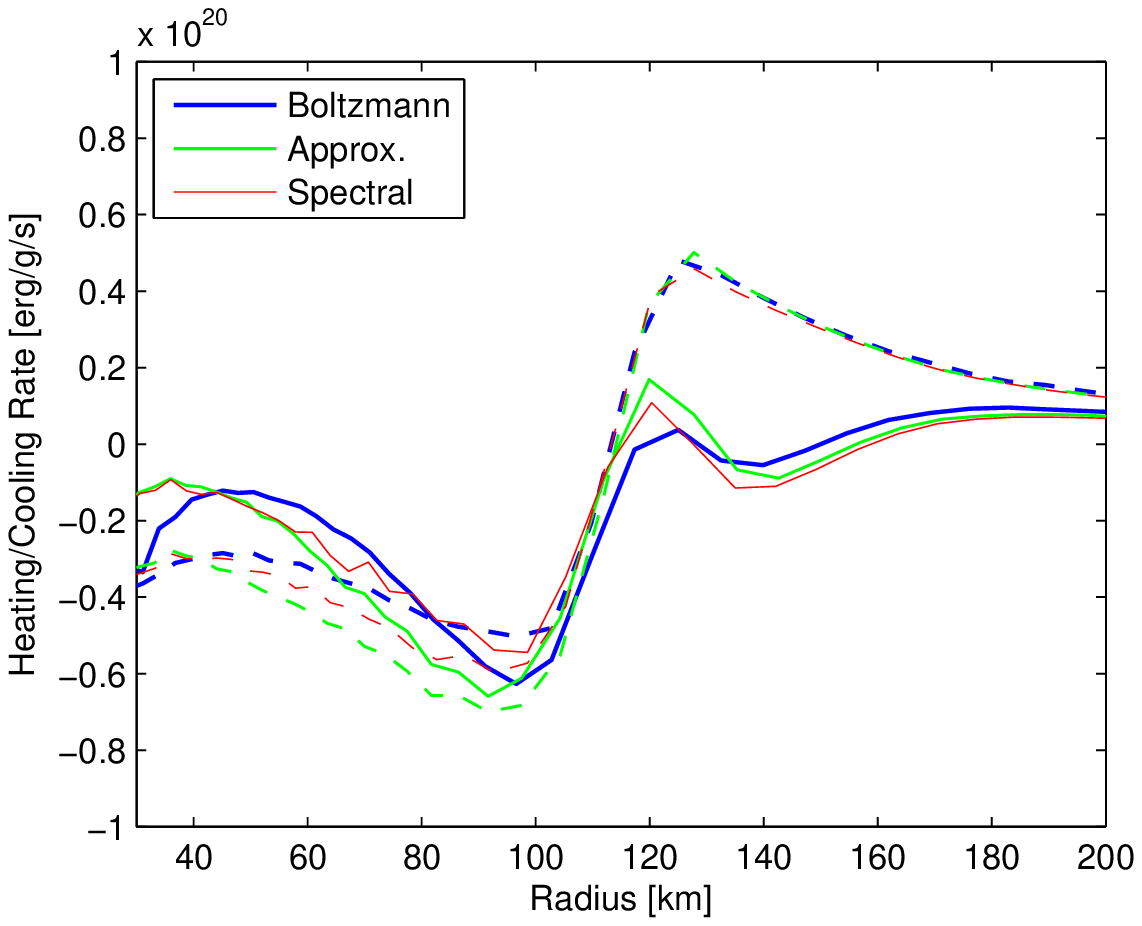}} \par}
\caption{Comparison of the heating or cooling rates specific to each neutrino
type at \protect\( 150 \) ms after bounce. The solid lines show the energy exchange
between neutrinos and matter.
The dashed lines show the energy exchange between antineutrinos and matter.
The lines labelled 'Approx.' and 'Spectral' have been calculated using the IDSA
with the standard and spectral representations of the trapped particle component,
respectively.}
\label{fig:fig6.ps}
\end{figure}
Once the trapped and streaming particle components are specified we
investigate the resulting net heating and cooling rates from neutrino absorption
and emission.  Fig. \ref{fig:fig6.ps} shows the energy exchange between neutrinos
and matter separately for the neutrinos (solid lines) and antineutrinos (dashed
lines). This comparison has to be performed and interpreted very carefully, because
the matter conditions are partially in equilibrium with the transported particles.
Even minor deviations from the stationary-state equilibrium at a restart of
the calculation may lead to large transient corrections in the exchange rates.
Hence, we let the hydrodynamics, the reference simulation and the IDSA solution
evolve after restart at \( 150 \) ms postbounce until the stationary-state is achieved
in both cases. The agreement of the IDSA (lines labelled with 'Approx.' and 'Spectra')
with the reference simulation (line labelled with 'Boltzmann') is excellent in the
heating region at radii larger than \( 120 \) km.
Differences of up to \( 30 \)\% can be seen in the cooling rates at smaller
radii in the rates for the antineutrinos (dashed lines). The IDSA with a spectral
representation of the trapped particles (lines labelled with 'Spectral') performs
better than the IDSA with the approximate representation of trapped particles
(lines labelled with 'Approx.').
The reference simulation shows stronger cooling by neutrinos than antineutrinos
at \( 95 \) km radius. This detail feature is not reproduced by either IDSA version.
However, as the cooling rates are a response to an intricate dynamical equilibrium
between the hydrodynamics and the quantities evolved by the transport algorithm,
a simple comparison of the heating and cooling rates alone is not sufficient to predict the
accuracy of a long-term evolution. In analogy, the instantaneous corrections applied
to the steering wheel of a car can vary from driver to driver, but most drivers follow
the trajectory of the road on the long term.

\begin{figure}
{\centering \resizebox*{\columnfigurewidth}{!}{\includegraphics{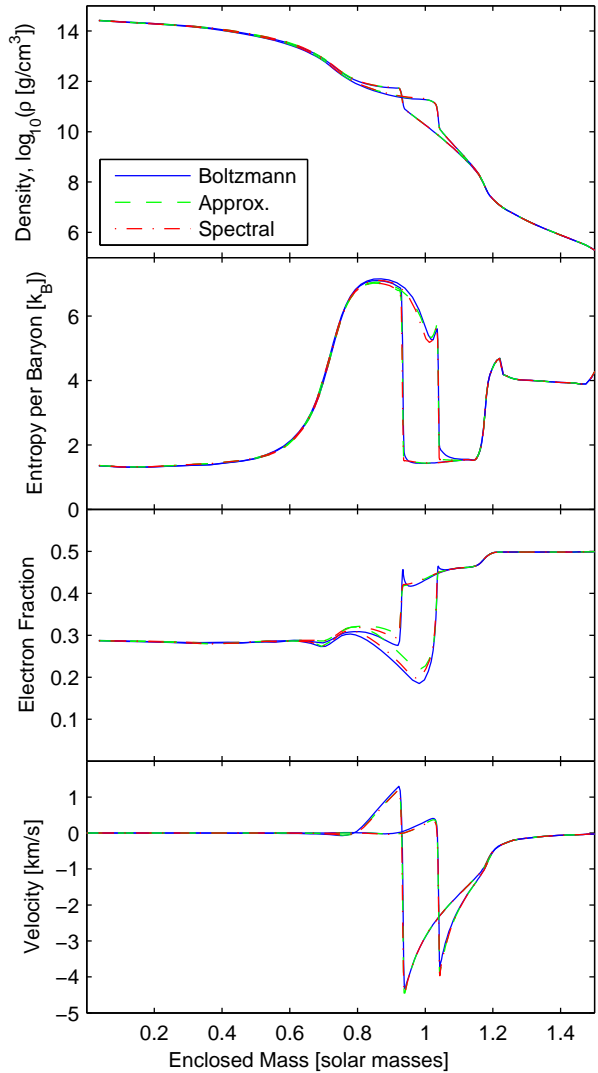}} \par}
\caption{Density, entropy, electron fraction and velocity as functions of enclosed
mass for a model based on Boltzmann neutrino transport (solid lines) and two
models based on the IDSA. The dashed lines represent data from the IDSA using
a thermal trapped particle distribution function while the dash-dotted lines represent
data from the IDSA with a spectral trapped particle treatment. The comparison is
shown at two different time instances: At
\protect\( 1\protect \) ms after bounce (lines with the shock discontinuity
positioned between \protect\( 0.8\protect \) and \protect\( 1\protect \)
M\protect\( _{\odot }\protect \)) and at \protect\( 3\protect \)
ms after bounce (lines with the shock discontinuity positioned between
\protect\( 1\protect \) and \protect\( 1.2\protect \) M\protect\( _{\odot }\protect \)).
Significant differences are only visible in the electron fraction profiles: The
deleptonisation in the approximate model is slightly delayed.
\label{fig:fig7.ps} }
\end{figure}
A comparison of profiles at the very early postbounce times of \( 1 \)
and \( 3 \) ms is shown in Fig. \ref{fig:fig7.ps}. No significant
deleptonisation has occured before \( 1 \) ms after bounce, while
at \( 3 \) ms after bounce the launch of the neutrino burst is reflected
in a trough of the electron fraction profile and a strong decline
of the entropy profile at an enclosed mass \( \sim 1 \) M\( _{\odot } \).
Significant differences between the IDSA and
the Boltzmann solution are only visible in the electron
fraction profiles.

\begin{figure}
{\centering \resizebox*{\columnfigurewidth}{!}{\includegraphics{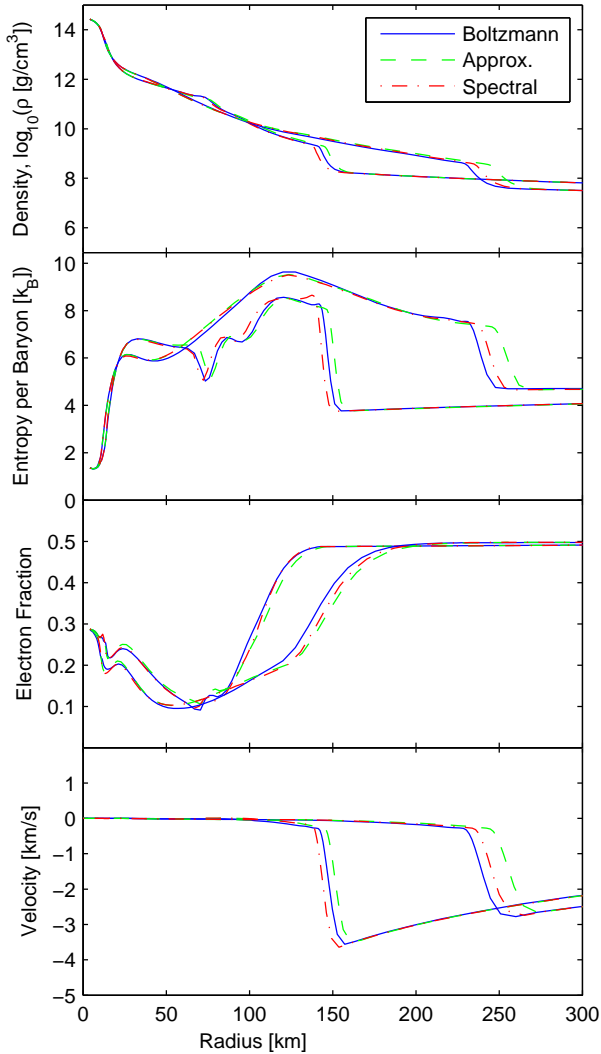}} \par}
\caption{Same presentation of the data as in Fig. \protect\ref{fig:fig7.ps}.
Here, the comparison is shown at two later time instances: At
\protect\( 30\protect \) ms after bounce (lines with the shock discontinuity
positioned at \protect\( 150\protect \) km radius) and at \protect\( 100\protect \)
ms after bounce (lines with the shock discontinuity positioned at
\protect\( 250\protect \) km radius). The two IDSA solutions lead to a somewhat
more optimistic shock expansion in this phase, but produce nicely detailed
features in the electron fraction and entropy profiles.
\label{fig:fig8.ps}}
\end{figure}

In fact, if one defines the diffusion source as described in Eq. (\ref{eq:definition.sigma.2}),
both IDSA versions lead to an earlier and somewhat faster deleptonisation
than in the Boltzmann solution. Our stationary-state assumption of
the free streaming particle component implies that produced neutrinos
propagate infinitely fast. This assumption is not applicable in the
very early postbounce phase because the dynamical changes of the matter
conditions occur on a similar time scale as the neutrino transport.
This transient deviation in the deleptonisation rates during the first
few milliseconds lets the models start on a different footing so that
the performance of the approximation in the more interesting later
postbounce phases becomes difficult to assess. In order to account
for the neutrino propagation time in an averaged way we therefore
limit the diffusion source \( \Sigma  \) in Eq. (\ref{eq:definition.sigma.2})
additionally by \( fc/\Delta r \), where \( \Delta r \) is an empirically
determined constant parameter that stands for a typical distance the
neutrinos have to propagate until stationary-state conditions become
applicable. The value \( \Delta r = 15 \) km leads to satisfactory
results. With this, the IDSA version with the spectral treatment of the
trapped particle distribution function (dash-dotted lines) leads to nice
agreement with the Boltzmann results (solid lines) while the IDSA version
with the two-parameter description of a thermal distribution function
(dashed lines) initially deleptonises slightly slower, but excellent
agreement is achieved later during the shock expansion phase. Corresponding
profiles at \( 30 \) ms and \( 100 \) ms after bounce are shown in Fig. \ref{fig:fig8.ps}.
The solutions based on the IDSA lead to a slightly more optimistic shock
expansion, but the detailed features in the entropy and electron fraction
profiles are accurately reproduced.

\begin{figure}
{\centering \resizebox*{\columnfigurewidth}{!}{\includegraphics{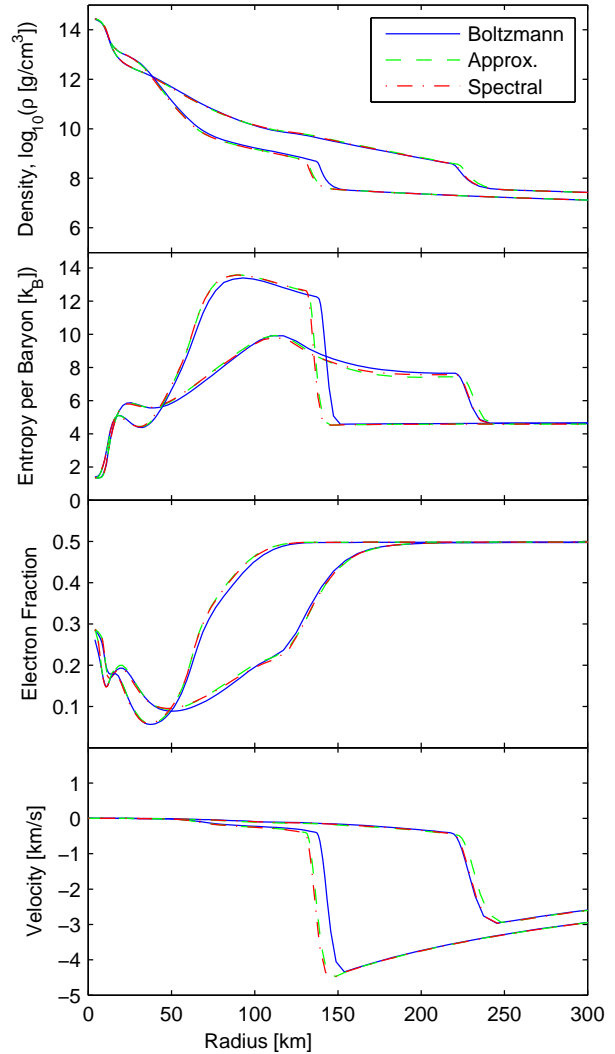}} \par}
\caption{Same presentation of the data as in Fig. \protect\ref{fig:fig7.ps}.
Here, the comparison is shown at two later time instances: At
\protect\( 150\protect \) ms after bounce (lines with the shock discontinuity
positioned at \protect\( 230\protect \) km radius) and at \protect\( 300\protect \)
ms after bounce (lines with the shock discontinuity positioned at
\protect\( 150\protect \) km radius). The evolution in the diffusive
domains is in good agreement, but the shock in the approximative
model retracts somewhat faster than in the accurate model. With respect
to a possible delayed explosion, the approximation produces the more pessimistic
model.\label{fig:fig9.ps}}
\end{figure}
The shock retraction phase is shown in Fig. \ref{fig:fig9.ps}. Shown
are profiles at \( 150 \) ms and \( 300 \) ms after bounce (note
that now the larger shock radius belongs to the earlier time slice).
The effect of neutrino heating (and hence successful neutrino transport) is
clearly visible in the entropy profile. Models that implement only neutrino leakage
develop a much smaller entropy that is entirely set by the heat dissipation at the
shock front. In the shock retraction phase, the peak entropy is obtained at the shock
front. If neutrino heating is switched on, the region of peak heating is located between
the neutrinospheres and the shock radius (see Fig. \ref{fig:fig6.ps}), because the neutrino
flux dilutes at larger distances and the matter density decreases in the outer layers.
Figs. \ref{fig:fig8.ps} and \ref{fig:fig9.ps} show the negative entropy gradient
between \( 120 \) km radius and the shock position that is caused by this neutrino
heating. As before we obtain excellent agreement in the
diffusive regime. But the model with neutrino transport approximations
starts to retract slightly earlier than the more accurate model with
Boltzmann neutrino transport. This is consistent with the larger cooling
rates of the IDSA model found in Fig. \ref{fig:fig6.ps}. However, the overall
differences are not much larger than differences that were also found in the first
comparison of two independent implementations of Boltzmann neutrino transport
\citep{Liebendoerfer.Rampp.ea:2005}. 
The chances of obtaining a delayed explosion seem to be more pessimistic in
the model with approximate neutrino transport.

\begin{figure*}
{\centering \resizebox*{\twocolumnfigurewidth}{!}{\includegraphics{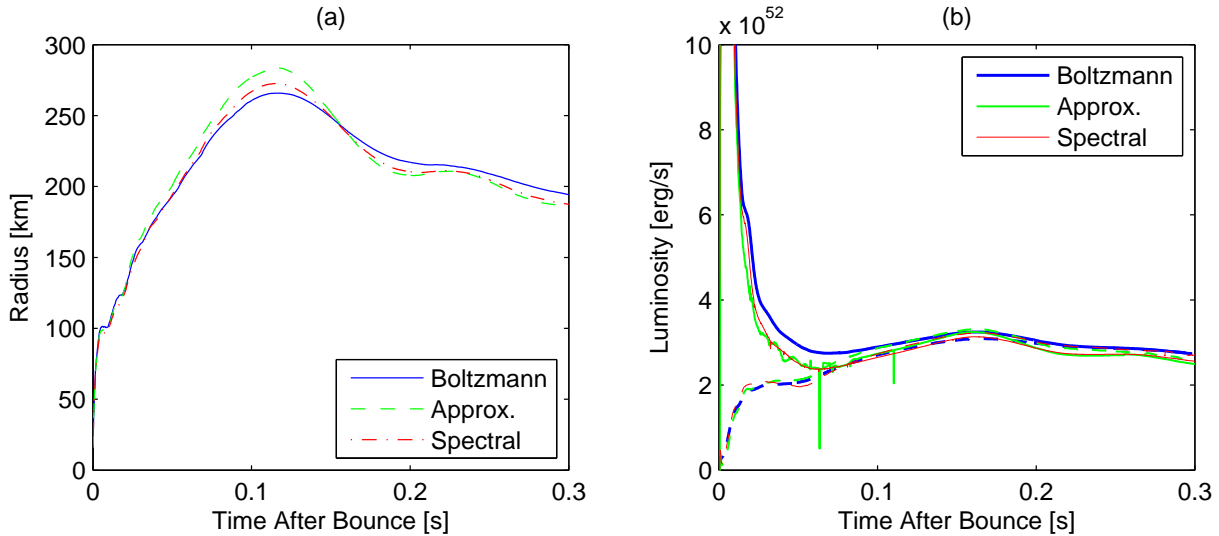}} \par}
\caption{Panel (a) shows the shock position as a function of time for a model
based on Boltzmann neutrino transport (solid line) and the two models using
the IDSA. Both approximate models agree
reasonably well with the reference model. The model based on a spectral
treatment of the trapped particle distribution function (dash-dotted line) is
somewhat closer than the standard IDSA approach (dashed line), but shows
the same qualitative deviations of a too optimistic initial expansion and a faster
shock retraction afterwards.
Panel (b) shows the neutrino luminosities for the
same three models. The agreement is satisfactory in general, the neutrino
luminosities (thin solid lines) are generally smaller in the approximate model than
in the Boltzmann model (thick solid line). The dashed lines refer to the antineutrino
luminosities. Some experienced numerical difficulties with
the approximate neutrino luminosites are discussed in the text.
\label{fig:fig10.ps}}
\end{figure*}
A short overview of the comparison is given in Fig. \ref{fig:fig10.ps}
in the form of shock trajectories and neutrino luminosities for the
three models. Panel (a) shows the shock trajectories. As stated before,
the agreement is excellent from the early postbounce phase through
the shock expansion phase. The models with approximate neutrino
transport initially produce a larger peak shock radius, but then compensate
by a faster shock retraction. The model with the spectral treatment of the
trapped particle component (dash-dotted line) is somewhat closer to the
reference simulation (solid line) than the standard IDSA with the two-parameter
description of the trapped particle distribution function (dashed line).
Panel (b) shows the neutrino luminosities measured in the comoving frame (this distinction
is only made for the Boltzmann reference results) at \( 500 \) km
radius. Overall, the luminosities are in good agreement. The neutrino
luminosities in the approximate models are somewhat smaller than
in the more accurate Boltzmann model.

While experimenting with different
variations in the approximation scheme we found that the stationary-state
luminosity in the approximate model is not a perfectly
smooth function of time, which leads us to the discussion of some
numerical issues related to the isotropic diffusion source approach.
The numerical details of our current implementation are outlined in
Appendix B. All local reaction rates
and the corresponding updates of the lepton fractions and temperature
are implemented in a time-implicit way. The non-local contribution
from the diffusion is also unconditionally stable because the updates
are ordered in such a way that each zone can use the updated distribution
function of its neighbour zone in the upwind direction (with respect
to the diffusion flux), while the contribution of the local distribution
function to the diffusion term is included in the time-implicit update.
However, the coupling to the stationary-state solution of the streaming
particle component is operator split. It is therefore still necessary to
restrict the time step to obtain a numerically stable evolution. In the diffusion
limit, the operator splitting is numerically stable due to above-mentioned
implicit finite differencing. In the free streaming limit, the operator splitting
is numerically stable due to the absence of relevant interactions between
the streaming particle component and matter. It is again the semi-transparent
transition regime where the operator splitting has the highest potential for
numerical instabilities. We obtain converged solutions if the time step chosen is
smaller than \( 1\) km\(/c\). This corresponds to the Courant-Friedrich-Levy
(CFL) limit for the speed of light or the time step limit of an explicitly finite
differenced diffusion scheme where the mean free path is half the zone
width, i.e. in the semi-transparent regime.
 If the time step chosen is larger, the stationary-state solution for the
streaming particle component starts to jump from time step to time step.
Rare luminosity jumps are still visible in Fig.
\ref{fig:fig10.ps}b. All other variables, e.g. the lepton fractions
and the temperature, do not react to these instantaneous luminosity
jumps because they evolve on a slower time scale where the luminosity
oscillations enter in a time-averaged manner.

In summary, we believe that the separation of particles into trapped
and streaming components provides an interesting ansatz for specifically
tailored simplifications of the Boltzmann transport equations. The
results in Figs. \ref{fig:fig7.ps}-\ref{fig:fig10.ps} show that such
a simplified descriptijon can reproduce the key features of neutrino
transport in supernova models. In the diffusion limit, the results
are stable and accurate. In the transition regime the time step
of our current implementation is at present limited by numerical fluctuations
of the stationary-state streaming particle flux at the transition to
transparent conditions. With our not thoroughly optimized code,
a single time step with the IDSA is of order \( 100 \) times faster than
a corresponding time step with Boltzmann neutrino transport. As the
fully implicit Boltzmann transport can take about \( 10 \) times larger
time steps, a full simulation with the IDSA is about an order of magnitude
faster than the one with Boltzmann transport. However, this number may vary from
application to application because the performance of the time-implicit Boltzmann
solution does not scale favourably with an increased dimensionality and size
of the computational domain. In three dimensions, we believe that simulations
based on the IDSA are feasible on average size high-performance computer
clusters, while simulations with comprehensive Boltzmann transport have not
yet been reported to be feasible even on top-performing computer systems.
A possible extension of the IDSA to multi-dimensional applications is outlined
in the following section.

\section{Generalization for multi-dimensional applications}

\label{sec:multid}So far, we have discussed the IDSA only in spherical symmetry.
As mentioned above, it is not the goal
of this paper to develop yet another approach that works only for
spherically symmetric models. Spherical symmetry is much better treated
with three-flavour Boltzmann transport where the comprehensive equations
of radiative transfer can consistently be solved \citep{Rampp.Janka:2002,Thompson.Burrows.Pinto:2003,Liebendoerfer.Messer.ea:2004,Sumiyoshi.Yamada.ea:2005}.
Hence, in this section we discuss how the scheme extends to the three-dimensional
case.

The state vector \( U \) of a 3D simulation only differs from Eq.
(\ref{eq:primitive.variables}) by the velocity \( v \), which becomes
a vector \( \vec{v}=\left( v_{x},v_{y},v_{z}\right)  \). With the
corresponding three directional components \( i=1\ldots 3 \) in the
momentum equation, the state vector \( U \) and flux vector \( F \)
become:\begin{equation}
\label{eq:3d.primitive.variables}
U=\left( \begin{array}{c}
\rho \\
\rho v_{i}\\
\rho \left( e+\frac{1}{2}v^{2}\right) \\
\rho Y_{{\rm e}}\\
\rho Y_{l}^{{\rm t}}\\
\left( \rho Z_{l}^{{\rm t}}\right) ^{\frac{3}{4}}
\end{array}\right) ,\quad F=\left( \begin{array}{c}
\vec{v}\rho \\
\vec{v}\rho v_{i}+p\\
\vec{v}\rho \left( e+\frac{1}{2}v^{2}+\frac{p}{\rho }\right) \\
\vec{v}\rho Y_{{\rm e}}\\
\vec{v}\rho Y_{l}^{{\rm t}}\\
\vec{v}\left( \rho Z_{l}^{{\rm t}}\right) ^{\frac{3}{4}}
\end{array}\right) .
\end{equation}
 This state vector can be evolved by a standard hydrodynamics scheme
that solves the multi-dimensional conservation law\begin{equation}
\label{eq:3d.conservation.law}
\frac{\partial }{\partial t}U+\vec{\nabla }\cdot \vec{F}=0.
\end{equation}

A more difficult part in the multi-dimensional IDSA is the consistent
solution of Eqs. (\ref{eq:definition.sigma.2})
and (\ref{eq:trapped.update.f})-(\ref{eq:trapped.update.e}). The
difficulty arises from the non-local scalar \begin{equation}
\label{eq:3d.alpha}
\alpha =\vec{\nabla }\cdot \left( \frac{-1}{3\left( j+\chi +\phi \right) }\vec{\nabla }f^{{\rm t}}\right) 
\end{equation}
in Eq. (\ref{eq:definition.sigma.2}). The other equations are local
and do not depend on the dimensionality of the problem, except that
the integration over the angle cosine, \( 1/2\int d\mu  \), must
be replaced by an integration over the entire solid angle of a sphere,
\( 1/\left( 4\pi \right) \int d\Omega  \).

We believe that there are three options when implementing Eq. (\ref{eq:3d.alpha}):
(i) One may calculate Eq. (\ref{eq:3d.alpha}) based on \( f^{{\rm t}} \)
of the previous time step and update \( f^{{\rm t}} \) locally with Eqs.
(\ref{eq:trapped.update.f})-(\ref{eq:trapped.update.e}). Hence,
only the local reactions would be implemented implicitly while the
diffusion part remains explicit. This approach is only conditionally
stable and the time step must be restricted to \( \Delta x^{2}/\left( 2\lambda c\right)  \).
The relevance of trapped particles diminishes in the transparent regime
around \( \lambda \sim \Delta x \). Thus, the time step restriction
is of similar order of magnitude as the CFL condition for the particle
propagation speed \( c \). As the sound speed in neutron stars is
close to the speed of light, the explicit approach may still be a
viable option for our application. (ii) In order to obtain an unconditionally
stable time step, a global implicit solution of above equations could
also be attempted. Because of the source limiting in Eq. (\ref{eq:definition.sigma.2}),
the problem is most easily expressed in terms of the unlimited diffusion
source \( \Sigma  \) as function of \( \partial f^{{\rm t}}/\partial t \)
as outlined in Appendix B in Eqs. (\ref{eq:fd.sigma.1}) and (\ref{eq:fd.implicit.dfdt}).
Then, one may search for a globally consistent solution of the field
of diffusion sources. This approach has the disadvantage that a spectral
problem has to be solved globally, which results in very memory- and
cpu-intensive code. The computational effort would become comparable
with multi-group flux limited diffusion or even more sophisticated
algorithms of radiative transfer. However, the solution could still
be restricted to the regimes where the trapped particle component
is dominant. (iii) Perhaps the most promising approach is the combination
of specific directional sweeps with a locally implicit update of the
diffusion source, be it in the form of a Crank-Nicholson scheme with
ADI (e.g. \citet{Press.Teukolsky.ea:1992}) or using Saul'yev's asymmetric
updates \citep{Saul'yev:1964,Tavakoli.Davami:2007}. This approach
leads to an unconditionally stable time step while retaining the efficiency
of an explicit scheme. In Appendix B we apply such an approch for
our spherically symmetric application. 

Independent of the method used, the result of this step determines
a partial update of the compositional quantities in \( U \) and a
spectral diffusion source \( \Sigma _{l}\left( E\right)  \) for each
particle species \( l \) at each grid point. This information is
stored for the following updates. The stationary-state solution for
the streaming particle component is then based on\begin{eqnarray}
\Delta \psi  & = & \frac{1}{4\pi }\int \left[ -\left( \hat{j}+\hat{\chi }\right) \hat{f}^{{\rm s}}+\hat{\Sigma }\right] d\hat{\Omega },\label{eq:3d.poisson} 
\end{eqnarray}
which is the natural extension of Eq. (\ref{eq:spherical.poisson})
to the multi-dimensional case. The integration over \( d\hat{\Omega } \)
is again performed over the solid angle of a sphere. If the relative
velocities in the transparent regime are large, the source should
be transformed from the comoving frame to the laboratory frame (quantities
measured in the laboratory frame carry a hat). Note that Eq. (\ref{eq:3d.poisson})
must be solved for each energy group and particle species separately.
But all directional information of the streaming particle flux \( \vec{\nabla }\psi  \)
can be retrieved from the corresponding scalar potential \( \psi _{l}( \hat{E})  \).
Once \( \psi _{l}( \hat{E})  \) has been calculated, the
equally large field \( \Sigma _{l}\left( E\right)  \) is no longer
needed.

After the hydrodynamic update according to Eq. (\ref{eq:3d.conservation.law}),
we have to convert the particle flux, \( \vec{\nabla }\psi _{l}( \hat{E})  \),
into a streaming particle density. This is now more ambiguous
than in Eq. (\ref{eq:lorentz.streaming}). One may still follow the
same procedure and first determine the particle scattering spheres.
Then, one assumes a flux factor of one half in regions enclosed by
the scattering spheres while using an isotropic emission ansatz along
the scattering spheres to estimate the flux factor in the domain outside
the scattering spheres based on their geometry. For nearly spherical
problems such as ours, the much simpler spherically symmetric ansatz described
above might even be sufficient to obtain an estimate of the flux factor.
Once the spectral streaming particle
density \( 1/\left( 4\pi \right) \int f_{l}^{{\rm s}}\left( E\right) d\Omega  \)
has been determined by the quotient of \( \vec{\nabla }\psi _{l}( \hat{E})  \)
and the flux factor, one cycle of updates in a multidimensional application is concluded.

\section{Conclusion}

The isotropic diffusion source approximation (IDSA) has been designed
to treat radiative transfer most efficiently in astrophysical applications
where particles diffuse out of an opaque domain that is subject to
multi-dimensional dynamics. This scenario occurs frequently in new-born
gravitationally bound objects, accretion discs, or dynamical atmospheres.
We have developed and tested the IDSA in the context
of spherically symmetric supernova models. The neutrino transport
in core collapse supernovae provides a very challenging application
where comprehensive solutions of the Boltzmann transport equation
have been studied in great detail (e.g. \citet{Mezzacappa:2005} and
references therein). Our approximation treats only the most important
features of radiative transfer. In a supernova model, these are (i) the
thermodynamics of trapped neutrinos (for example the \( dE_{\nu} = -p_{\nu}dV \)-term),
(ii) the correct diffusion limit,
(iii) the spectrum of transported neutrinos, and (iv) the angular focussing
of the neutrino propagation directions with increasing distance from
the neutrinospheres.

We implemented the IDSA in the hydrodynamics
code Agile \citep{Liebendoerfer.Rosswog.Thielemann:2002}
to compare its performance with a more sophisticated transport model.
We compare the postbounce evolution of a \( 13 \) M\( _{\odot } \)
star \citep{Nomoto.Hashimoto:1988} in Newtonian gravity with the corresponding
model N13 that is based on Boltzmann neutrino transport
\citep{Liebendoerfer.Rampp.ea:2005}. We find good
agreement between the results of the IDSA and the reference
model during the early postbounce phase when the neutrino burst is
launched and the accretion shock expands to its maximum radius. The
IDSA leads to a somewhat larger maximum shock radius and a faster
shock retraction thereafter, i.e. a slightly more pessimistic model with respect
to the possibility of a delayed supernova explosion. We conclude that the
concept of the IDSA accurately captures the feedback of neutrino transport
onto the hydrodynamics evolution in spherically symmetric supernova
models. The neutrino fluxes and spectra are in nice agreement if small time
steps are taken, but unphysical fluctuations in the stationary-state neutrino
luminosities develop when the time steps are increased much beyond
the neutrino propagation time scale. We will continue to improve the
numerical stability of the scheme in future work. 

With respect to the choice of physical assumptions and with respect
to the quality of the results, the IDSA is most closely
related to the well-known flux-limited diffusion (FLD) approximation.
Like the latter, the IDSA is guided by the diffusion
limit at high opacities and the free streaming limit at low opacities.
The FLD approximation
is based on diffusive fluxes that are limited by the physical particle
speed at the transition to the free streaming regime. This limiting is
well-defined in a spherically symmetric setup where the limiting can
be applied to the radial velocity direction. In a multi-dimensional simulation,
however, FLD fails to determine the correct direction of the net flux outside
the diffusive regime. In the free streaming regime there is no physical
connection between the local radiation intensity and the flux direction.
A related point is the efficiency of the FLD approach. Since the implemented
numerical solution strategies are optimised for the diffusive domain,
they easily become inefficient in the free streaming regime, which
represents often the larger part of the computational domain. These are
the two main shortcomings of FLD which we address with the IDSA.
The IDSA is based on an overlapping two-component description. The trapped
particle component dominates in the diffusive regime while the streaming
component dominates in the free streaming regime. In the diffusive regime,
the IDSA is equivalent to FLD. In the semi-transparent regime, the IDSA
is based on a different interpolation between the diffusive and the free
streaming regimes. In the IDSA, the limiter is applied to the sources
instead of the fluxes. In the free streaming limit of a multi-dimensional
application, the direction of fluxes is well-defined by the non-local position
of the sources. The distinction between the trapped and streaming particle components
allows us to separately test more efficient solution algorithms for both regimes.

For example, we assumed that the trapped particle component is close to
thermal and weak equilibrium while the streaming particle component
carries a non-equilibrium spectrum. In the framework of the IDSA we
have compare a grey and a spectral (i.e. energy-averaged and energy-dependent)
treatment of the evolution of  the trapped particle component.
We found visible differences of several percent. We believe
that the differences are large enough to value the spectral approach, but at
the same time they are small enough to be ignored if the efficiency of the
algorithm in 3D simulations becomes critical. However, we emphasise that
this comparison must not be confused with the comparison of a globally grey
scheme and a globally spectral approach. We believe that the {\em streaming} particle
component, which represents the lepton and energy flux in a supernova
application, must be treated in a spectral manner. The location of the
neutrinospheres is highly energy-dependent. Hence, the geometry of the
radiative transfer problem is different for each neutrino energy. An energy-averaged
local quantity can only be expected to be accurately represented if one averages over
solutions of the energy-dependent transport problem. A grey scheme, that solves an
energy-averaged transport problem instead, is unlikely to produce accurate results
in the supernova context.

Aside from the field of neutrino-driven supernova models, FLD has been used
in fields ranging from planet
migration \citep{Paardekooper.Mellema:2006} to cosmology \citep{Bonanno.Romano:1994},
including areas such as accretion discs, both around stars \citep{Kley:1989}
and more energetic objects where the accretion disc is dominated by
radiation (e.g. \citep{Turner.Stone.Sano:2002,Hujeirat.Camenzind:2000}),
and star formation \citep{Whitehouse.Bate:2006}. The IDSA is potentially
applicable to these and other
areas of astrophysics where FLD is commonly used. However, neither
the IDSA nor the FLDA can replace the development of more reliable
transport schemes because, for each application, it has to be ascertained
that no important physical ingredients are missed. We hope that the
IDSA is well-suited to accompany the application of sophisticated
radiative transfer algorithms in order to enable an efficient exploration
of input physics, parameter space, or dimensionality in computer models
where a comprehensive solution of the Boltzmann transport equation
is not affordable.

\section*{Acknowledgements}

This research was funded by the Swiss National Science Foundation
under grant No. PP002-106627/1. ML and SCW acknowledge the hospitality
of the Institute for Pure and Applied Mathematics at the University
of California, Los Angeles, where our collaboration began. We are
grateful for prolific discussions with Ue-Li Pen, Tom Abel, Jose
Pons, Yudai Suwa and Andrey Yudin.

\section*{Appendix A: The diffusion limit revisited}

\label{sec:appendix.diffusion.limit}Even if the following is old
wine in a new hose, we derive the diffusion limit here in brief in
order to make a clear distinction between a flux-limited diffusion
approximation, which determines the \emph{particle flux} based on
the diffusion limit and our scheme, which requires a well-founded
approximation for the \emph{collision integral} in the diffusive limit.

The source term on the right hand side of the Boltzmann equation (\ref{eq:boltzmann}) features
an isotropic emissivity, \( j \), an isotropic absorptivity, \( \chi \), and an isoenergetic
scattering kernel, \( R \). For the derivation of the diffusion limit, the angular dependence of
the scattering kernel is usually expanded in  a Legendre series (see e.g.
\citet{Bruenn:1985,Mezzacappa.Bruenn:1993a}),
\begin{equation}
\label{eq:legendre.expansion}
\frac{E^{2}}{c\left( hc\right) ^{3}} R\left( E,\mu ,\mu^{\prime}\right)
= \frac{1}{4\pi}\sum_{\ell}
\left( 2\ell +1\right) \phi_{\ell} \left( E\right)  \int_0^{2\pi} P_{\ell} \left( \cos\theta \right) d\varphi ,
\end{equation}
where \( \cos\theta = \mu\mu^{\prime}
+ \cos\varphi \left[ \left( 1-\mu^2 \right)
\left( 1-\mu^{\prime 2} \right) \right]^{1/2} \)
specifies the angle between the incoming and the scattered particle propagation
directions. In Eq. (\ref{eq:legendre.expansion}) we define \( \phi_{\ell} \left( E\right) \) such
that it enters Eq. (\ref{eq:boltzmann}) without additional factors.
If we denote the operator on the left hand side of
Eq. (\ref{eq:boltzmann}) by \( D\left( f\right)  \) and truncate the Legendre expansion
after the second term, we reach a very concise
representation of the Boltzmann equation,
\begin{equation}
\label{eq:boltzmann.compact}
D\left( f\right) =j-\left( j+\chi +\phi_0 \right) f
+ \phi_0 \frac{1}{2}\int _{-1}^{+1}f d\mu
+ 3\mu\phi_1\frac{1}{2}\int_{-1}^{+1}f \mu d\mu.
\end{equation}
From the zeroth and first angular moments of Eq. (\ref{eq:boltzmann.compact}) we extract an
expression for the zeroth and first moments of the distribution function, respectively,
\begin{eqnarray}
\frac{1}{2}\int f d\mu &=& \frac{1}{j+\chi} \left( j - \frac{1}{2}\int D\left( f\right) d\mu \right),\nonumber
\label{eq:zeroth.first.moment}\\
\frac{1}{2}\int f \mu d\mu &=& \frac{-1}{j+\chi+\phi_0-\phi_1}\frac{1}{2}\int D\left( f\right) \mu d\mu.
\end{eqnarray}
If we substitute these terms back into Eq. (\ref{eq:boltzmann.compact}),  we can express
the distribution function \( f\) in the following way,
\begin{equation}
\label{eq:f.as.function.of.Df}
f = \frac{1}{j+\chi+\phi_0}
\left\{ j-D\left( f\right)
+ \frac{\phi_0}{j+\chi}\left[ j - \frac{1}{2}\int D\left( f\right) d\mu \right]
- \frac{\mu\phi_1}{j+\chi+\phi_0-\phi_1} \frac{1}{2}\int D\left( f\right) \mu d\mu
\right\}.
\end{equation}

Now, we perform a Chapman-Enskog expansion \citep{Chapman.Cowling:1970}.
To this purpose we introduce
a small expansion parameter, \( \varepsilon  \), and replace the large
source functions \( j \), \( \chi  \) and \( \phi_{0,1}  \) in Eq. (\ref{eq:f.as.function.of.Df})
by \( \bar{j}/\varepsilon  \), \( \bar{\chi }/\varepsilon  \) and \( \bar{\phi}_{0,1}/\varepsilon  \),
respectively. From the zeroth order terms in \( \varepsilon \) we then obtain as expected
the isotropic equilibrium distribution function
\( f_0 = j/\left( j+\chi \right) \). However, we can also determine the first order corrections
to the equilibrium distribution function:
\begin{equation}
\varepsilon f_1 = \frac{-1}{j+\chi+\phi_0}
\left[ D\left( f_0\right)
+ \frac{\phi_0}{j+\chi} \frac{1}{2}\int D\left( f_0\right) d\mu
+ \frac{\mu\phi_1}{j+\chi+\phi_0-\phi_1} \frac{1}{2}\int D\left( f_0\right) \mu d\mu
\right].
\end{equation}
After these preparations, we obtain the leading order particle number exchange rate with
matter, \( s \), by substituting the approximate distribution
function \( f\simeq f_0+\varepsilon f_1 \) into Eq. (\ref{eq:boltzmann.compact}) and integrating
the left hand side over particle propagation angles:
\begin{equation}
\label{eq:net.source}
s = \frac{1}{2} \int D\left( f_0 + \varepsilon f_1 \right) d\mu.
\end{equation}

This integral can be simplified if one decomposes the operator \( D\left( f \right) \)
into an operator \( D^{+}\left( f\right)  \)
containing terms that are symmetric in \( \mu  \) and an operator
\( D^{-}\left( f\right)  \) containing terms that are antisymmetric
in \( \mu  \). Hence, \( D\left( f\right) =D^{+}\left( f\right) +D^{-}\left( f\right)  \),
with
\begin{eqnarray}
D^{+}\left( f\right)  & = & \frac{df}{cdt}+
\left( \frac{\partial \ln \rho }{c\partial t}+\frac{3v}{cr}\right)
\mu \left( 1-\mu ^{2}\right) \frac{\partial f}{\partial \mu }
+ \left[ \mu ^{2}\left( \frac{\partial \ln \rho }{c\partial t}
+\frac{3v}{cr}\right) -\frac{v}{cr}\right] E\frac{\partial f}{\partial E}\nonumber \\
D^{-}\left( f\right)  & = & \mu \frac{\partial f}{\partial r}+\frac{1}{r}\left( 1-\mu ^{2}\right) \frac{\partial f}{\partial \mu }.\label{eq:dplus.dminus} 
\end{eqnarray}
Only the symmetric terms \( D^{+}\left( f_{0}\right)  \), \( D^{+}\left[ D^{+}\left( f_{0}\right) \right]  \)
and \( D^{-}\left[ D^{-}\left( f_{0}\right) \right]  \) survive the
angular integration in Eq. (\ref{eq:net.source}). Furthermore,
we may neglect the \( D^{+}\left[ D^{+}\left( f\right) \right]  \)
terms because they are of higher order in the expansion than the leading
\( 1/2\int D^{+}\left( f_{0}\right) d\mu  \) term. The \( D^{-}\left[ D^{-}\left( f_{0}\right) \right]  \)
terms, however, are of leading order because there is no lower
order contribution involving \( D^{-} \). The intermediate result
\begin{equation}
\frac{1}{2}\int D\left( \varepsilon f_1\right) d\mu =
\frac{1}{2}\int D^{-} \left(\frac{-1}{j+\chi+\phi_0} \left[ D^{-}\left( f_0 \right)
+ \frac{\mu\phi_1}{j+\chi+\phi_0-\phi_1}
\frac{1}{2}\int D^{-} \left( f_0\right)\mu d\mu \right]\right)
\end{equation}
can be further simplified if one explicitly calculates the integral in the last term using
\( D^{-} \left( f_0 \right) = \mu\partial f_0 /\partial r \). This leaves us with
the following leading order source terms in the diffusion limit,
\begin{equation}
s=\frac{1}{2}\int D^{+}\left( f_{0}\right) d\mu -\frac{1}{2}\int D^{-}
\left( \frac{1}{j+\chi +\phi_0 - \phi_1 }D^{-}\left( f_{0}\right) \right) d\mu .
\end{equation}
 The original operators in Eq. (\ref{eq:dplus.dminus}) 
 may now be readily substituted. Several \( \partial /\partial \mu  \)-terms drop out
due to the isotropy of \( f_{0}=j/\left( j+\chi \right)  \) and the
remaining angular integrations can easily be performed. With this,
we obtain the following simple expression for the collision integral
in the diffusion limit, \begin{equation}
\label{eq:diffusion}
s=\frac{df_{0}}{cdt}+\frac{1}{3}\frac{\partial \ln \rho }{c\partial t}E\frac{\partial f_{0}}{\partial E}-\frac{1}{r^{2}}\frac{\partial }{\partial r}\left( \frac{r^{2}}{3\left( j+\chi +\phi\right) }\frac{\partial f_{0}}{\partial r}\right),
\end{equation}
where, in order to shorten the notation, we define \( \phi = \phi_0 - \phi_1 \).
Note that a flux-limited diffusion scheme applies Eq. (\ref{eq:diffusion})
as evolution equation for an unknown \( f\left( t,r,E\right)  \)
(instead of \( f_{0} \)) together with a source function \( s \left( f\right) \).
Our scheme uses Eq. (\ref{eq:diffusion}) as leading order approximation
to the source \( s \) for a given equilibrium distribution function
\( f_{0} \), exactly as derived. The first term in Eq. (\ref{eq:diffusion})
describes sources that arise from adjustments to a time-dependent
equilibrium, the second term accounts for changes in the average particle
energy when the fluid is compressed or expanded and the third term
describes the divergence of the diffusive flux in the rest frame of
the fluid according to the transport mean free path \( \lambda _{t}=\left( j+\chi +\phi \right) ^{-1} \).

\section*{Appendix B: Implementation and Finite Differencing}

\label{sec:appendix.implementation} In this appendix we discuss the
finite differencing of our implementation of the IDSA. Our scheme solves
Eq. (\ref{eq:trapped.update.f})
for the trapped particles, Eq. (\ref{eq:spherical.poisson}) for the
streaming particles, and Eq. (\ref{eq:conservation.law}) for the
hydrodynamics in an operator splitting manner. The spherically symmetric
computational domain is represented by an array of concentric fluid
shells labelled by the index \( i=1\ldots i_{max} \). Unprimed indices
are used for zone centre values and primed indices for the upper zone edge
values. Energy-dependent quantities are calculated for
a set of discrete particle energies \( E_{k} \), where \( k=1\ldots k_{max} \).
This representation exists for several time instances,
which we label by an upper index \( n=1\ldots n_{max} \). For example, \(
\lambda_{i',k}^{n} \) refers to the mean free path \( \lambda \left( r_{i+1/2},E_{k},t^{n} \right) \)
at the zone edge between \( r_{i} \) and \( r_{i+1} \), at energy \( E_{k} \) and at
time \( t^{n} \).

A time steps begins with the construction of the particle flux at time \( t^{n+1} \)
from the particle sources, \( \Sigma - \left( j+\chi\right)\frac{1}{2}\int f^{{\rm s}} d\mu \), which
have been calculated and stored during the previous time step \( t^{n} \).
The particle flux is given by the gradient of the potential \( \psi \) as described in
Eq. (\ref{eq:spherical.poisson}). In spherical symmetry, this Poisson equation
is easily solved by the integration of the sources over the volume,
\begin{equation}
\label{eq:fd.poisson}
\left( \frac{\partial \psi }{\partial r}\right) _{i',k}^{n+1}=\frac{1}{r_{i'}^{2}}\sum _{j=1}^{j=i'}
\left[ \Sigma _{j,k}^{n}-\tilde{\chi }_{j,k}^{n}\frac{1}{2}\int f_{j,k}^{{\rm s},n}d\mu \right] r^{2}_{j}\left( r_{j'}-r_{j'-1}\right),
\end{equation}
where we used a tilde to denote stimulated
absorption \( \tilde\chi = j+\chi \). At present we neglect the Lorentz transformation of the
sources from the comoving frame to the laboratory frame.

Next, we retrieve the current conditions from the solution
vector \( U_{i}^{n}=U(r_{i},t^{n}) \) and calculate the energy-dependent transport mean
free path \( \lambda _{i,k}^{n} \) and the optical depth
\( \tau _{i,k}^{n} =\int _{r_{i}}^{\infty}\left( 1/\lambda (r)_{k}^{n}\right) dr \).
In spherical symmetry,
the energy-dependent radius \( R_{\nu ,k}^{n} \) of the neutrinosphere
is found at the point where \( \tau \left( R_{\nu ,k}^{n}\right) =2/3 \).
The neutrinosphere is then used to derive the streaming particle
density from the streaming particle flux according to Eq. (\ref{eq:lorentz.streaming}).
Again, we neglect the Lorentz transformation between the laboratory and
comoving frames and approximate the streaming particle density by
\begin{equation}
\frac{1}{2}\int f_{i,k}^{{\rm s},n+1}d\mu  = \frac{2}{1+\sqrt{1-\left[ R^{n}_{\nu ,k}/\max \left( r,R_{\nu ,k}^{n}\right) \right] ^{2}}}
\left( \frac{r_{i'-1}}{r_{i}}\right) ^{2}\max \left[ \left( \frac{\partial \psi }{\partial r}\right) _{i'-1,k}^{n+1},0\right] .\label{eq:fd.streaming.density} 
\end{equation}
The factor \( \left( r_{i'-1}/r_{i}\right) ^{2} \) converts the flux
from the inner zone edge \( i'-1 \), where it has been calculated
in Eq. (\ref{eq:fd.poisson}), to the zone centre \( r_{i} \), where
it is used in the calculation of the diffusion source. Two measures
significantly increase the stability of the scheme. On the one hand
it is important to exclude the contribution of zone \( i \) to the
streaming particle flux. On the other hand, the maximum function with the square brackets
eliminates fluxes that stream against the density gradient.

The trapped particle distribution functions are separated into a thermal component,
and a spectral perturbation,
\begin{equation}
f_{i,k}^{{\rm t},n} = f_{k}^{{\rm th}}\left( Y_{i}^{{\rm t},n}, Z_{i}^{{\rm t},n} \right) + \delta f_{i,k}^{{\rm t},n}.
\label{eq:definition.distft}
\end{equation}
The thermal part is represented by a Fermi function, \( f_{k}^{{\rm th}} \), where the
particle temperature and particle degeneracy are chosen such that
Eq. (\ref{eq:definition.yt.zt}) is fulfilled for the current values of the quantities
\( Y_{i}^{{\rm t},n} \) and \( Z_{i}^{{\rm t},n} \) in the solution vector \( U_{i}^{n} \).
If the analytical approximations given in \citep{Epstein.Pethick:1981} are
used, a consistent set of the particle temperature and degeneracy parameters
can be obtained from the solution of a one-dimensional Newton-Raphson scheme.
For the standard version of the IDSA we neglect the perturbations by setting
\( \delta f_{i,k}^{{\rm t},n} \equiv 0 \). In order to test this approximation in Sect.
\ref{sec:verification}, we implemented a second version of the IDSA, where the
deviation of the trapped particle spectrum from the thermal spectrum is evolved
in time as described in Eq. (\ref{eq:spectral.correction}).

The next step is the search for a consistent solution
of Eqs. (\ref{eq:trapped.update.f}), (\ref{eq:trapped.update.ye})
and (\ref{eq:trapped.update.e}). In most of the following monochromatic equations,
all quantities carry
the same energy index \( k \). To further reduce the amount of indices
we will write the energy index only if quantities at different energies
are involved in one equation. We discretise the fast
reaction rates on the right hand side of Eq. (\ref{eq:trapped.update.f})
in a time-implicit way,\begin{equation}
\label{eq:fd.explicit.dfdt}
\frac{f_{i}^{{\rm t},n+1}-f_{i}^{{\rm t},n}}{c\Delta t}=j_{i}^{n+1}-\tilde{\chi }_{i}^{n+1}f_{i}^{{\rm t},n+1}-\Sigma _{i}^{n+1},
\end{equation}
and eliminate the dependence
on \( f_{i}^{{\rm t},n+1} \) on the right hand side by rewriting Eq. (\ref{eq:fd.explicit.dfdt})
in the form\begin{equation}
\label{eq:fd.implicit.dfdt}
\frac{f_{i}^{{\rm t},n+1}-f_{i}^{{\rm t},n}}{c\Delta t}=\frac{j^{n+1}_{i}-\tilde{\chi }_{i}^{n+1}f_{i}^{{\rm t},n}-\Sigma _{i}^{n+1}}{1+\tilde{\chi }_{i}^{n+1}c\Delta t}.
\end{equation}
The diffusion source \( \Sigma  \), which is defined in Eq. (\ref{eq:definition.sigma.2}),
depends on \( \alpha  \). We will work with the following definition
of \( \alpha  \): \begin{eqnarray}
\xi _{i} & = & \frac{1}{3r_{i}^{2}\left( r_{i'}-r_{i'-1}\right) }\frac{r_{i'}^{2}\lambda _{i'}}{r_{i+1}-r_{i}}\label{eq:fd.xi} \\
\zeta _{i} & = & \frac{1}{3r_{i}^{2}\left( r_{i'}-r_{i'-1}\right) }\frac{r_{i'-1}^{2}\lambda _{i'-1}}{r_{i}-r_{i-1}}\label{eq:fd.zeta} \\
\eta _{i} & = & \xi _{i}+\zeta _{i}\label{eq:fd.eta} \\
\alpha _{i} & = & -\xi _{i}f_{i+1}^{{\rm t}}+\eta _{i}f_{i}^{{\rm t}}-\zeta _{i}f_{i-1}^{{\rm t}}.\label{eq:fd.alpha} 
\end{eqnarray}

In the search for a consistent diffusion source we first assume that
\( \tilde{\Sigma }=\alpha +\tilde{\chi }\frac{1}{2}\int f^{{\rm s}}d\mu  \)
is not limited. Then we apply the limiters in Eq. (\ref{eq:definition.sigma.2})
to the result. It is known that the numerical evolution of the diffusion
equation is not unconditionally stable unless the term \( \alpha  \)
is finite differenced in an at least partially implicit manner (e.g.
\citet{Press.Teukolsky.ea:1992}). Hence, we would like to discretise
the diffusion source as
\begin{equation}
\tilde{\Sigma }_{i}^{n+1} = -\xi _{i}^{n}\left( f_{i+1}^{{\rm t},n+1}-f_{i+1}^{{\rm t},n}\right) +\eta _{i}^{n}\left( f_{i}^{{\rm t},n+1}-f_{i}^{{\rm t},n}\right) -\zeta _{i}^{n}\left( f_{i-1}^{{\rm t},n+1}-f_{i-1}^{{\rm t},n}\right)
+ \alpha _{i}^{n}+\tilde{\chi }_{i}^{n+1}\frac{1}{2}\int f_{i}^{{\rm s},n+1}d\mu .
\label{eq:fd.sigma.1} 
\end{equation}
The dependence of this equation on the unknown distribution functions
\( f^{{\rm t},n+1} \) can be removed if we substitute Eq. (\ref{eq:fd.implicit.dfdt})
in the first three terms. The result is a non-local equation involving
\( \tilde{\Sigma }_{i+1}^{n+1} \), \( \tilde{\Sigma }_{i}^{n+1} \) and \( \tilde{\Sigma }_{i-1}^{n+1} \),
which could globally be solved for \( \tilde{\Sigma }_{i}^{n+1} \). However,
since our target application is based on a three-dimensional parallelised
hydrodynamics code, for the moment we try to avoid a global solution
of Eqs. (\ref{eq:fd.sigma.1}) and (\ref{eq:fd.implicit.dfdt}). At
least in our spherically symmetric example, the diffusive fluxes propagate
almost exclusively outward. Hence we finite difference the outward propagating
flux implicitly and the inward propagating flux explicitly, which leads to
\begin{equation}
\tilde{\Sigma }_{i}^{n+1} = \zeta _{i}^{n}\left( f_{i}^{{\rm t},n+1}-f_{i}^{{\rm t},n}\right) -\zeta _{i}^{n}\left( f_{i-1}^{{\rm t},n+1}-f_{i-1}^{{\rm t},n}\right)
+ \alpha _{i}^{n}+\tilde{\chi }_{i}^{n+1}\frac{1}{2}\int f_{i}^{{\rm s},n+1}d\mu .
\label{eq:fd.sigma.2} 
\end{equation}
If one applies Eq. (\ref{eq:fd.sigma.2}) from the inside out,
\( f_{i-1}^{{\rm t},n+1} \) is known from the update of the previous zone. After substitution of Eq.
(\ref{eq:fd.implicit.dfdt}) in the first term, Eq. (\ref{eq:fd.sigma.2}) can be solved for
\( \tilde\Sigma _{i}^{n+1} \) so that the limiter from Eq. (\ref{eq:definition.sigma.2}) can
be applied,
\begin{eqnarray}
\Sigma _{i}^{n+1} & = & \min \left\{ \max \left[ \tilde{\Sigma }_{i}^{n+1},0\right] ,j_{i}^{n+1}\right\} \nonumber \\
\tilde{\Sigma }_{i}^{n+1} & = & \frac{1}{1+\left( \zeta _{i}^{n}+\tilde{\chi }_{i}^{n+1}\right) c\Delta t}\left\{ \zeta _{i}^{n}c\Delta t\left( j_{i}^{n+1}-\tilde{\chi }_{i}^{n+1}f_{i}^{{\rm t},n}\right) +\left( 1+\tilde{\chi }_{i}^{n+1}c\Delta t\right) \right. \nonumber \\
 & \times  & \left. \left[ -\xi _{i}^{n}f_{i+1}^{{\rm t},n}+\eta _{i}^{n}f_{i}^{{\rm t},n}-\zeta _{i}^{n}f_{i-1}^{{\rm t},n+1}+\tilde{\chi }_{i}^{n+1}\frac{1}{2}\int f_{i}^{{\rm s},n+1}d\mu \right] \right\} .
\label{eq:fd.sigma.3} 
\end{eqnarray}
 This diffusion source \( \Sigma _{i}^{n+1} \) can now be used in Eq. (\ref{eq:fd.implicit.dfdt})
to determine the discrete time derivative of the trapped particle
distribution function, \( \left( f_{i}^{{\rm t},n+1}-f_{i}^{{\rm t},n}\right) /\left( c\Delta t\right)  \).
Furthermore, this derivative specifies the net interaction rate between
matter and trapped particles, which can be calculated according to
Eq. (\ref{eq:trapped.update.matter}),\begin{equation}
\label{eq:fd.s}
s_{i}^{n+1}=\frac{f_{i}^{{\rm t},n+1}-f_{i}^{{\rm t},n}}{c\Delta t}+\Sigma _{i}^{n+1}-\tilde{\chi }_{i}^{n+1}\frac{1}{2}\int f_{i}^{{\rm s},n+1}d\mu .
\end{equation}

In the application of Eq. (\ref{eq:fd.sigma.3}) to the supernova
model in Sect. \ref{sec:verification} we observed a slightly too rapid
deleptonisation few milliseconds after the bounce of the stellar core.
As the dynamical time scale of the bounce is comparable to the neutrino
propagation time scale in this short transition phase, the stationary-state
assumption of our approximation may not hold. This effect disappeared
when we additionally limited \( \Sigma \) in Eq. (\ref{eq:fd.sigma.3})
by \( f^{{\rm t},n+1}/\Delta r \),
where \( \Delta r\sim 15 \) km is an empirically determined constant
parameter. 

In the case of neutrino transport with electron neutrinos and electron
antineutrinos Eq. (\ref{eq:fd.s}) leads to the following 
updates of the electron fraction and the internal specific
energy in Eqs. (\ref{eq:trapped.update.ye}) and (\ref{eq:trapped.update.e}):\begin{eqnarray}
\frac{Y_{{\rm e},i}^{n+1}-Y_{{\rm e},i}^{n}}{c\Delta t}+\frac{m_{\rm b}}{\rho^{n} _{i}}\frac{4\pi }{\left( hc\right) ^{3}}\sum _{k}\left( s_{\nu _{{\rm e}},i,k}^{n+1}-s_{\bar{\nu}_{{\rm e}},i,k}^{n+1}\right) E_{k}^{2}dE_{k} & = & 0\label{eq:fd.ye} \\
\frac{e_{i}^{n+1}-e_{i}^{n}}{c\Delta t}+\frac{m_{\rm b}}{\rho^{n} _{i}}\frac{4\pi }{\left( hc\right) ^{3}}\sum _{k}\left( s_{ \nu _{\rm e},i,k}^{n+1}+s_{\bar{\nu }_{\rm e},i,k}^{n+1}\right) E_{k}^{3}dE_{k} & = & 0.\label{eq:fd.e} 
\end{eqnarray}
However, the neutrino-matter interactions \( j_{i}^{n+1}=j\left( \rho^{n} _{i},Y_{{\rm e},i}^{n+1},e_{i}^{n+1}\right)  \)
and \( \chi \left( \rho^{n} _{i},Y_{{\rm e},i}^{n+1},e_{i}^{n+1}\right)  \),
which enter the right hand sides of Eqs. (\ref{eq:fd.ye}) and (\ref{eq:fd.e})
via Eq. (\ref{eq:fd.implicit.dfdt}), depend on the updated values
for the electron fraction and specific internal energy. Hence we solve
this nonlinear algebraic system of equations numerically by two-dimensional
Newton-Raphson iterations to determine a consistent \( \left\{ Y_{{\rm e},i}^{n+1},e_{i}^{n+1}\right\}  \)-pair%
\footnote{In fact, for compatibility reasons with the hydrodynamics code and
the equation of state, our implementation of the Newton-Raphson scheme
works with the logarithmic entropy instead of the specific energy.
}. This consistency is important to accurately represent reactive equilibria
at high opacities. Following an approach of \citet{Mezzacappa.Messer:1999},
we first evaluate Eq. (\ref{eq:fd.s}) on four corners of a rectangle,
centred around the initial values \( \left\{ Y_{{\rm e},i}^{n},e_{i}^{n}\right\}  \)
in the space spanned by electron fraction and specific energy. Then,
we use interpolations in this rectangle to efficiently obtain the
reaction rates and their derivatives with respect to \( Y_{{\rm e}} \)
and \( e \) during the iterations required to find a consistent solution
of Eqs. (\ref{eq:fd.ye}) and (\ref{eq:fd.e}). Once the iterations
have converged, we update \( \rho Y_{{\rm e}} \) and \( \rho e \) in the
hydrodynamics state vector \( U^{n'} \), where \( n' \) indicates
an intermediate incomplete update.

Finally, the components \( \rho Y^{{\rm t}} \) and \( \left( \rho Z^{{\rm t}}\right) ^{3/4} \) in \( U^{n} \)
are updated based on Eqs. (\ref{eq:trapped.update.yt}), (\ref{eq:trapped.update.zt})
and (\ref{eq:fd.implicit.dfdt}). In the standard version of the IDSA, the spectral information
of the trapped particle component is discarded at this point. For the alternative
version, the spectral information is stored in an energy-dependent
vector \( \delta f_{i,k}^{{\rm t},n+1} \) that will be used in the next time step according
to Eq. (\ref{eq:definition.distft}). The latter is obtained from
\begin{eqnarray}
\delta f_{k}^{*} & = & f_{i,k}^{{\rm t},n+1} - f_{k}^{{\rm th}}\left( Y_{i}^{{\rm t},n+1}, Z_{i}^{{\rm t},n+1} \right) \nonumber\\
\delta f_{i,k}^{{\rm t},n+1} & = & \delta f_{k}^{*}
+ \frac{a_{0}b_{2} - a_{1}b_{1} - \left( a_{0}b_{1} - a_{1}b_{0} \right) E_{k}}
{\left( b_{1}^2 - b_{0}b_{2} \right)} f_{k}^{{\rm th}}\left( Y_{i}^{{\rm t},n+1}, Z_{i}^{{\rm t},n+1} \right)\label{eq:spectral.correction}\\
a_{m} & = & \sum_{k} \delta f_{k}^{*} E_{k}^{2+m} dE_{k} \nonumber\\
b_{m} & = & \sum_{k} f_{k}^{{\rm th}}\left( Y_{i}^{{\rm t},n+1}, Z_{i}^{{\rm t},n+1} \right) E_{k}^{2+m} dE_{k}. \nonumber
\end{eqnarray}
The first line in Eq. (\ref{eq:spectral.correction}) calculates the deviation between the
updated trapped particle distribution function and the thermal spectrum. Due to numerical
inaccuracies, spurious errors can accumulate in \( \delta f_{k}^{*} \) over many time steps
and start to contribute to the trapped particle abundance \( Y_{i}^{{\rm t}} \)
and the mean specific energy \( Z_{i}^{{\rm t}} \) which should be determined by the thermal
part alone. Adding the correction term on the second line of Eq. (\ref{eq:spectral.correction})
guarantees that
\( \sum \delta f_{i,k}^{{\rm t},n+1} E_{k}^{2} dE_{k} = 0 \) and
\( \sum \delta f_{i,k}^{{\rm t},n+1} E_{k}^{3} dE_{k} = 0 \) to machine precision. The correction
term depends on energy moments of \( \delta f_{k}^{*} \) and the thermal distribution
function. 

The cycle of updates is concluded by evolving the partially updated
hydrodynamics state \( U^{n'} \) to \( U^{n+1} \) according to Eq.
(\ref{eq:conservation.law}) (and including gravity) by any standard
hydrodynamics scheme. For this demonstration of the method we use
the spherically symmetric hydrodynamics code Agile
\citep{Liebendoerfer.Rosswog.Thielemann:2002} with a second order
accurate TVD advection scheme.
\end{document}